\documentstyle[twoside,amsfonts,epsfig]{article}

\catcode`\@=11
\long\def\@makefntext#1{
\protect\noindent \hbox to 3.2pt {\hskip-.9pt  
$^{{\eightrm\@thefnmark}}$\hfil}#1\hfill}		%CAN BE USED 

\def\thefootnote{\fnsymbol{footnote}}
\def\@makefnmark{\hbox to 0pt{$^{\@thefnmark}$\hss}}	%ORIGINAL 
	
\def\ps@myheadings{\let\@mkboth\@gobbletwo
\def\@oddhead{\hbox{}
\rightmark\hfil\eightrm\thepage}   
\def\@oddfoot{}\def\@evenhead{\eightrm\thepage\hfil
\leftmark\hbox{}}\def\@evenfoot{}
\def\sectionmark##1{}\def\subsectionmark##1{}}

\oddsidemargin=\evensidemargin
\renewcommand{\thefootnote}{\fnsymbol{footnote}}

\newcounter{sectionc}\newcounter{subsectionc}\newcounter{subsubsectionc}
\renewcommand{\section}[1] {\vspace{12pt}\addtocounter{sectionc}{1} 
\setcounter{subsectionc}{0}\setcounter{subsubsectionc}{0}\noindent 
	{\tenbf\thesectionc. #1}\par\vspace{5pt}
\def\artun_section{\thesectionc}
}

\renewcommand{\subsection}[1] {\vspace{12pt}\addtocounter{subsectionc}{1} 
	\setcounter{subsubsectionc}{0}\noindent 
	{\bf\thesectionc.\thesubsectionc. {\kern1pt \bfit
	#1}}\par\vspace{5pt}
\def\artun_section{\thesectionc.\thesubsectionc}
}
\renewcommand{\subsubsection}[1] {\vspace{12pt}\addtocounter{subsubsectionc}{1}
	\noindent{\tenrm\thesectionc.\thesubsectionc.\thesubsubsectionc.
	{\kern1pt \tenit #1}}\par\vspace{5pt}
\def\artun_section{\thesectionc.\thesubsectionc.\thesubsubsectionc}
}
\newcommand{\nonumsection}[1] {\vspace{12pt}\noindent{\tenbf #1}
	\par\vspace{5pt}}

\newcounter{appendixc}
\newcounter{subappendixc}[appendixc]
\newcounter{subsubappendixc}[subappendixc]
\renewcommand{\thesubappendixc}{\Alph{appendixc}.\arabic{subappendixc}}
\renewcommand{\thesubsubappendixc}
	{\Alph{appendixc}.\arabic{subappendixc}.\arabic{subsubappendixc}}

\renewcommand{\appendix}[1] {\vspace{12pt}
        \refstepcounter{appendixc}
        \setcounter{figure}{0}
        \setcounter{table}{0}
        \setcounter{lemma}{0}
        \setcounter{theorem}{0}
        \setcounter{corollary}{0}
        \setcounter{definition}{0}
        \setcounter{equation}{0}
        \renewcommand{\thefigure}{\Alph{appendixc}.\arabic{figure}}
        \renewcommand{\thetable}{\Alph{appendixc}.\arabic{table}}
        \renewcommand{\theappendixc}{\Alph{appendixc}}
        \renewcommand{\thelemma}{\Alph{appendixc}.\arabic{lemma}}
        \renewcommand{\thetheorem}{\Alph{appendixc}.\arabic{theorem}}
        \renewcommand{\thedefinition}{\Alph{appendixc}.\arabic{definition}}
        \renewcommand{\thecorollary}{\Alph{appendixc}.\arabic{corollary}}
        \renewcommand{\theequation}{\Alph{appendixc}.\arabic{equation}}
        \noindent{\tenbf Appendix \theappendixc #1}\par\vspace{5pt}}
\newcommand{\subappendix}[1] {\vspace{12pt}
        \refstepcounter{subappendixc}
        \noindent{\bf Appendix \thesubappendixc. {\kern1pt \bfit #1}}
	\par\vspace{5pt}}
\newcommand{\subsubappendix}[1] {\vspace{12pt}
        \refstepcounter{subsubappendixc}
        \noindent{\rm Appendix \thesubsubappendixc. {\kern1pt \tenit #1}}
	\par\vspace{5pt}}

\newcommand{\textlineskip}{\baselineskip=13pt}
\newcommand{\smalllineskip}{\baselineskip=10pt}

\def\eightcirc{
\begin{picture}(0,0)
\put(4.4,1.8){\circle{6.5}}
\end{picture}}
\def\eightcopyright{\eightcirc\kern2.7pt\hbox{\eightrm c}} 

\newcommand{\copyrightheading}[1]
	{\vspace*{-2.5cm}\smalllineskip{\flushleft
	 }}

\def\abstracts#1#2#3{{
	\centering{\begin{minipage}{4.5in}\baselineskip=10pt\footnotesize
	\parindent=0pt #1\par 
	\parindent=15pt #2\par
	\parindent=15pt #3
	\end{minipage}}\par}} 

\newcommand{\bibit}{\nineit}
\newcommand{\bibbf}{\ninebf}
\renewenvironment{thebibliography}[1]
	{\frenchspacing
	 \ninerm\baselineskip=11pt
	 \begin{list}{\arabic{enumi}.}
	{\usecounter{enumi}\setlength{\parsep}{0pt}
	 \setlength{\leftmargin 12.7pt}{\rightmargin 0pt} %FOR 1--9 ITEMS
	 \setlength{\itemsep}{0pt} \settowidth
	{\labelwidth}{#1.}\sloppy}}{\end{list}}

\newcounter{itemlistc}
\newcounter{romanlistc}
\newcounter{alphlistc}
\newcounter{arabiclistc}

\newcommand{\fcaption}[1]{
        \refstepcounter{figure}
        \setbox\@tempboxa = \hbox{\footnotesize Fig.~\thefigure. #1}
        \ifdim \wd\@tempboxa > 5in
           {\begin{center}
        \parbox{5in}{\footnotesize\smalllineskip Fig.~\thefigure. #1}
            \end{center}}
        \else
             {\begin{center}
             {\footnotesize Fig.~\thefigure. #1}
              \end{center}}
        \fi}

\newcommand{\tcaption}[1]{
        \refstepcounter{table}
        \setbox\@tempboxa = \hbox{\footnotesize Table~\thetable. #1}
        \ifdim \wd\@tempboxa > 5in
           {\begin{center}
        \parbox{5in}{\footnotesize\smalllineskip Table~\thetable. #1}
            \end{center}}
        \else
             {\begin{center}
             {\footnotesize Table~\thetable. #1}
              \end{center}}
        \fi}

\def\@citex[#1]#2{\if@filesw\immediate\write\@auxout
	{\string\citation{#2}}\fi
\def\@citea{}\@cite{\@for\@citeb:=#2\do
	{\@citea\def\@citea{,}\@ifundefined
	{b@\@citeb}{{\bf ?}\@warning
	{Citation `\@citeb' on page \thepage \space undefined}}
	{\csname b@\@citeb\endcsname}}}{#1}}

\newif\if@cghi
\def\cite{\@cghitrue\@ifnextchar [{\@tempswatrue
	\@citex}{\@tempswafalse\@citex[]}}
\def\citelow{\@cghifalse\@ifnextchar [{\@tempswatrue
	\@citex}{\@tempswafalse\@citex[]}}
\def\@cite#1#2{{$\null^{#1}$\if@tempswa\typeout
	{IJCGA warning: optional citation argument 
	ignored: `#2'} \fi}}
\def\pmb#1{\setbox0=\hbox{#1}
	\kern-.025em\copy0\kern-\wd0
	\kern.05em\copy0\kern-\wd0
	\kern-.025em\raise.0433em\box0}

\def\fnt#1#2{\footnotetext{\kern-.3em
	{$^{\mbox{\scriptsize #1}}$}{#2}}}

\def\fpage#1{\begingroup
\voffset=.3in
\thispagestyle{empty}\begin{table}[b]\centerline{\footnotesize #1}
	\end{table}\endgroup}

\def\runninghead#1#2{\pagestyle{myheadings}
\markboth{{\protect\footnotesize\it{\quad #1}}\hfill}
{\hfill{\protect\footnotesize\it{#2\quad}}}}
\font\tenrm=cmr10
\font\tenit=cmti10 
\font\tenbf=cmbx10
\font\bfit=cmbxti10 at 10pt
\font\ninerm=cmr9
\font\nineit=cmti9
\font\ninebf=cmbx9
\font\eightrm=cmr8

\def\qed{\hbox{${\vcenter{\vbox{			%HOLLOW SQUARE
   \hrule height 0.4pt\hbox{\vrule width 0.4pt height 6pt
   \kern5pt\vrule width 0.4pt}\hrule height 0.4pt}}}$}}

\renewcommand{\thefootnote}{\fnsymbol{footnote}}	%USE SYMBOLIC FOOTNOTE

\newcommand{\M}{{\cal M}}
\def\lsi{\raise0.3ex\hbox{$<$\kern-0.75em\raise-1.1ex\hbox{$\sim$}}}
\def\gsi{\raise0.3ex\hbox{$>$\kern-0.75em\raise-1.1ex\hbox{$\sim$}}}
\newcommand{\lsim}{\mathop{\lsi}}

\def\selabel#1{\@bsphack
  \protected@write\@auxout{}%
         {\string\newlabel{#1}{{\artun_section}{\thepage}}}%
  \@esphack}
\newcommand\ie{{\it i.e.}}

\def\acite#1{\if@filesw\immediate\write\@auxout
	{\string\citation{#1}}\fi
\def\@citea{}\@acite{\@for\@citeb:=#1\do
	{\@citea\def\@citea{,}\@ifundefined
	{b@\@citeb}{{\bf ?}\@warning
	{Citation `\@citeb' on page \thepage \space undefined}}
	{\csname b@\@citeb\endcsname}}}}
\def\@acite#1{#1}

\begin{document}

\runninghead{Formation of topological defects in gauge field theories}
{Formation of topological defects in gauge field theories}

\normalsize\textlineskip
\thispagestyle{empty}
\setcounter{page}{1}

\copyrightheading{}			%{Vol. 0, No. 0 (1993) 000--000}

\fpage{1}

\rightline{DAMTP-2001-78}
\vspace*{0.035truein}
\rightline{20 August, 2001}

\vspace*{0.68truein}

\centerline{\bf FORMATION OF TOPOLOGICAL DEFECTS}
\vspace*{0.035truein}
\centerline{\bf IN GAUGE FIELD THEORIES}
\vspace*{0.37truein}
\centerline{\footnotesize ARTTU RAJANTIE}
\vspace*{0.015truein}
\centerline{\footnotesize\it Department of Applied Mathematics and
Theoretical Physics}
\centerline{\footnotesize\it University of Cambridge, 
Wilberforce Road}
\centerline{\footnotesize\it Cambridge CB3 0WA, United Kingdom}
\centerline{\footnotesize\it E-mail: a.k.rajantie@damtp.cam.ac.uk}
\baselineskip=10pt
\vspace*{0.225truein}
%\publisher{(received date)}{(revised date)}

\vspace*{0.21truein}
\abstracts{
When a symmetry gets spontaneously broken in a phase transition,
topological defects
are typically formed.
The theoretical picture of how this happens in a breakdown of a global
symmetry, the Kibble-Zurek mechanism, 
is well established and has been tested in various
condensed matter experiments. 
However, from the viewpoint of particle physics and cosmology,
gauge field
theories are more relevant than global theories. In recent years,
there have been significant advances in the theory of defect formation
in gauge field theories, which make precise predictions possible,
and in experimental techniques that can be
used to test these predictions in superconductor experiments.
This opens up the possibility of carrying out relatively simple and
controlled experiments, in which the non-equilibrium phase transition
dynamics of gauge field theories can be studied. This will have 
a significant impact on our understanding of phase transitions in the
early universe and in heavy ion collider experiments.
In this paper,
I review the current status of the theory and the experiments
in which it can be
tested.
}{}{}

%\textlineskip			%) USE THIS MEASUREMENT WHEN THERE IS
%\vspace*{12pt}			%) NO SECTION HEADING

\vspace*{1pt}\textlineskip	%) USE THIS MEASUREMENT WHEN THERE IS
\section{Introduction}	%) A SECTION HEADING
\textheight=7.8truein
\setcounter{footnote}{0}
\renewcommand{\thefootnote}{\alph{footnote}}

\noindent
The formation of topological defects in phase transitions is a very generic
phenomenon in physics.
It can be studied experimentally in different condensed
matter systems, but it is also believed to have happened during the
early evolution of the universe,\cite{Kibble:1976sj,VilenkinShellard} 
and the produced topological defects
may have observable consequences to the properties of the universe
today. Defect formation also has many similarities with the
non-equilibrium phenomena that take place in heavy ion 
collisions.\cite{Berdnikov:2000ph,Boyanovsky:2001nt}

In this article, however, we discuss defect formation in field theories
as a physical phenomenon in its own right,
as a unique way of obtaining information about the non-equilibrium
dynamics of field theories. 
Theoretical calculations of phase transition dynamics are only
possible in practice if one is willing to make many simplifying
approximations and assumptions, and it is not at all clear which of
these are actually justified. In this respect, it is useful that
quantum field theories are applicable very generally in physics.
Very similar field theories describe a wide range of systems from
superconductors to interactions of elementary particles. In
particular, the problems faced in theoretical calculations are
identical
in these systems. This raises the hope of testing some
theoretical predictions in condensed matter
experiments,\cite{Zurek:1985qw}
in order to
find out which approximations are reliable and how one should treat
the field theories of particle physics theoretically.

To a certain extent, this hope has already been realized in equilibrium
field theory. Critical phenomena can be studied experimentally in many
condensed matter systems, and they form the basis for the
non-perturbative picture of quantum field theories.\cite{Zinn-Justin}
In particular, the
concept of renormalization group, which is based on the critical
behaviour of field theories near second order transitions points, is
crucial for our understanding of particle physics.

In the case of non-equilibrium dynamics, both experiments and
theoretical calculations are significantly
more difficult, and therefore less
progress has been made. In particular, it is very difficult to study
directly the non-equilibrium dynamics during a phase transition in any
experiment. Topological defects play a special role in this respect,
because they
survive for a long time after
the phase transition, when the system has otherwise equilibrated.
Their number density and spatial distribution carry information about
the dynamics of the phase transition, and they can be compared to
predictions of theoretical calculations.

In many ways, this approach is very similar to particle cosmology,
where we hope to extract some information about particle physics from
the relics of the Big Bang, and to particle accelerator experiments,
in
which
we cannot observe the collision itself, but by studying its
products we can reconstruct it. The difference is that superconductors
and other condensed matter systems are theoretically simpler to treat
and experimentally they allow much more control of different
parameters. Of course, they cannot give any direct information about
particle physics, but they will, nonetheless, 
help us understand the non-equilibrium
dynamics of gauge field theories and in particular how they
can be studied theoretically.

Indeed, the need for reliable tools for studying non-equilibrium gauge
field theories is now greater than ever before, as the 
Relativistic Heavy Ion Collider, which has already started operation,
and the Large Hadron Collider, which is scheduled to start by 2005,
will be able to study directly the quark-gluon
plasma phase of quantum chromodynamics. The heavy-ion collisions in
these experiments are so
complicated that very reliable and accurate theoretical
calculations are needed in order to confront the experimental
results, but our
 present understanding of the theory is too
rudimentary for that. The insight provided by condensed matter
experiments is therefore likely to be extremely useful.
In particular, it is believed that at a certain value of the beam
energy, the quark-gluon plasma produced in the collision cools
through a second-order transition point.\cite{Stephanov:1998dy} 
The
relevant dynamics can then be described as freezing out of
long-wavelength modes,\cite{Berdnikov:2000ph,Boyanovsky:2001nt} 
very much like in the theories of defect
formation.

So far, the formation of topological defects has been studied
in liquid crystal\cite{TurokNature,Bowick:1994rz} 
and superfluid experiments,\cite{Hendry,Bauerle,Ruutu:1996qz,Dodd}
which are systems with
global symmetries. The theoretical scenario that is believed to be
applicable in this case is the Kibble-Zurek 
mechanism,\cite{Kibble:1976sj,Zurek:1985qw}
but the experiments have produced mixed results.
From the viewpoint of particle physics and cosmology, theories with
global symmetries are less interesting than gauge field theories, 
whose natural condensed matter realization is
superconductivity. The dynamics of gauge field theories are 
more complicated, but
a theory of defect formation has been developed 
recently\cite{Hindmarsh:2000kd}
for Abelian systems, such as superconductors.
At the same time, the rapid progress in experimental techniques in the
recent years has made accurate superconductor experiments possible.

The purpose of this paper is to review the current theoretical status
of defect formation in field theories, with a special
emphasis on gauge field theories.
The various
applications of topological defects in other fields such as 
cosmology are outside the scope
of this article, and have been discussed in other 
reviews.\cite{VilenkinShellard,Hindmarsh:1995re,Magueijo:2000se}

Throughout this paper, we shall use natural units, \ie, $c=\hbar=k_B=\mu_0=1$.
The structure of this paper is the following:
In Section~\ref{sect:Topo}, we discuss the
classification and basic properties of topological defects in field theories.
In Section~\ref{sect:finiteT}, we present the basics of
finite-temperature field theories and phase transitions. Defect
formation in first-order phase transitions is discussed in
Section~\ref{sect:1storder}. Second-order phase transitions are more
complicated but also more relevant physically, and are
discussed in Section~\ref{sect:2ndorder}. In Section~\ref{sect:methods},
we present different approximations that have been used to study the
dynamics of phase transitions theoretically, and in
Section~\ref{sect:expt}, we discuss different condensed matter
experiments that have been carried out or proposed.

%\newpage
\section{Topological defects}
\selabel{sect:Topo}
\noindent

\subsection{Spontaneous symmetry breakdown}
\selabel{sect:SSB}
\noindent
In most field theories, 
the Lagrangian ${\cal L}[\phi]$ is invariant under certain 
transformations $\phi\rightarrow g(\phi)$, \ie,
\begin{equation}
{\cal L}[g(\phi)]={\cal L}[\phi],
\end{equation}
where $g$ is an element of
the symmetry group $G$, typically a Lie group. The fields of the theory,
which are here denoted by $\phi$ for simplicity, can be in any
representation of the group, and this determines the action of the
transformation $g$ on the field configuration. 

In particle physics, these symmetries are often {\em gauge} 
symmetries, which
means that the group element $g$ can be a function of space and time.
For instance, the Lagrangian of quantum electrodynamics is
\begin{equation}
{\cal L}=-\frac{1}{4}F_{\mu\nu}F^{\mu\nu}+
\overline{\psi}(i\gamma^\mu D_\mu-m)\psi,
\end{equation}
where $D_\mu=\partial_\mu+ieA_\mu$ and $F_{\mu\nu}=\partial_\mu
A_\nu-\partial_\nu A_\mu$ are the covariant derivative and the field
strength tensor, respectively.
This Lagrangian is invariant under U(1) gauge transformations
\begin{equation}
\psi(x)\rightarrow\exp(i\alpha(x))\psi(x),\qquad
A_\mu(x)\rightarrow A_\mu(x)-e^{-1}\partial_\mu\alpha(x).
\end{equation}
In the limit $e\rightarrow 0$, the gauge field $A_\mu$ decouples, and
the Lagrangian is only invariant if $\alpha$
is constant in space. In this case, the symmetry is said to be {\em global}.

Gauge symmetries are extremely important in particle physics, because
besides electromagnetism, weak and strong
interactions are described by SU(2) and SU(3) gauge groups,
respectively.
Global symmetries, on the other hand, 
are more common in condensed matter physics.

There are situations in which a symmetry of the Lagrangian is not
reflected by the state of the system, and the symmetry is
said to be {\it spontaneously broken}. Let us, for instance, consider a 
model with a single complex scalar field
\begin{equation}
\label{equ:phi4def}
{\cal L}=\partial_\mu\phi^*\partial^\mu\phi-V(|\phi|).
\end{equation}
This Lagrangian is invariant under global U(1) transformations
$\phi\rightarrow\exp(i\alpha)\phi$, where $\alpha$ is a constant in 
space-time. 
In a classical field theory, the vacuum state is simply the one in which
$\phi$ minimizes the potential $V$ and is constant in space and time.
If, for instance, the
potential has the form
\begin{equation}
V(|\phi|)=m^2|\phi|^2+\lambda|\phi|^4,
\end{equation}
and $m^2<0$,
there is a 
set of degenerate minima at
$|\phi|=(-m^2/2\lambda)^{1/2}\equiv v$, and therefore the system has many vacua, which are related
to each other by symmetry transformations. Small
perturbations
around any of these vacua do not possess the U(1) symmetry, which
means that the symmetry is broken. In the presence of quantum or
thermal fluctuations, the situation is more subtle, and we cannot
say whether the symmetries are broken simply by looking at the
shape of the 
potential. These cases are discussed in more detail in
Section~\ref{sect:finiteT}.

If the symmetry group $G$ that is broken is more complicated than
U(1), it may also be partially broken. This happens if the vacuum
remains invariant under a subgroup $H$ of $G$. Suppose, for
instance, that 
the full symmetry group is SU(2), and it is broken by an adjoint scalar field
$\Phi=\sum_i\Phi^i\sigma^i$, where $\sigma^i$ are the Pauli
matrices. Under symmetry transformations, $\Phi$ transforms as
$\Phi\rightarrow g^\dagger\Phi g$, where $g$ is an SU(2)
matrix. Because all vacuum states are equivalent, we can assume
$\Phi=\Phi^3\sigma^3$. Then $\Phi$ is
invariant under transformations $g=\exp(i\alpha\sigma^3)$, which
form a U(1) subgroup of the full SU(2) symmetry group.

\subsection{Classification of defects}
\noindent
Let us now consider static classical solutions when
a symmetry is spontaneously broken.
To be a solution, a field configuration must approach
the vacuum asymptotically at infinity. However, instead of a unique
vacuum, the system has a set $\M$
of vacua, which is called the vacuum manifold. In the above examples
of U(1) and SU(2) symmetries, $\M$ is topologically a circle and a
sphere, respectively.

It is possible to have field configurations in which $\phi$
approaches a different point in $\M$ in different directions.
In this case, the values of $\phi$ at spatial infinity form a mapping
$\phi: S^{D-1}\rightarrow \M$, where $D$ is the dimensionality of the
space. 
The homotopy classes of this mapping consist of configurations that
can be continuously transformed into one another and
form the $(D-1)$th 
homotopy group $\pi_{D-1}(\M)$ of $\M$. Furthermore, by restricting the
mapping to subsets $S^n$ of the $S^{D-1}$ at infinity, where $n<D-1$,
we obtain the lower homotopy groups $\pi_n(\M)$. 

Because the elements
of the homotopy groups are invariant under continuous deformations, a
field configuration that corresponds to a non-trivial element of any
$\pi_n(\M)$ with $n\le D-1$ cannot be continuously transformed into
the vacuum solution. 
Within each homotopy class, there is a field configuration with
minimal energy, and that configuration is necessarily a static
solution of the field equations. However, it has typically a higher
energy than the vacuum. In $D=3$ dimensions, configurations with
non-trivial $0$th, $1$st and $2$nd homotopies correspond to domain walls,
vortex lines (strings) and monopoles, respectively.

As an example, let us consider the U(1) symmetric theory
(\ref{equ:phi4def}) 
and a
field configuration of the form
$\phi(r,\varphi,z)=|\phi(r)|\exp(i\varphi)$ in cylindrical coordinates
where $\varphi$ is the azimuth angle.
A configuration like this cannot be continuously transformed into a
vacuum solution, because the path of $\phi$ at the infinity
corresponds to a non-trivial element of the first homotopy group
$\pi_1(\M)$. 
If we want $\phi$ to be continuous, it must vanish at $r=0$,
and this costs energy. 
Therefore we have a classical solution that corresponds to a
line-like object, a vortex line.

The topology of the vacuum manifold follows directly from the symmetry
breaking pattern.
Because different
vacua are characterized by different values of $\phi$, $\M$ is
homeomorphic to
the set $G(\phi)$ 
of all the possible values that can be obtained from $\phi$ by
symmetry transformations. However, since the transformations in the
unbroken subgroup $H$ leave $\phi$ unchanged, $\M$ is homeomorphic to
the coset space $G/H$ rather than the full group $G$. Thus, the
homotopy groups $\pi_n(G/H)$ tell what types of topological defects
exist in the theory.

In smooth configurations, which are not necessarily
solutions of the classical field
equations, we can identify the topological defects by calculating the
element of $\pi_n(\M)$ to which the behaviour of the field $\phi$
on a closed surface of dimensionality $n$ corresponds. If this winding
number is non-trivial, 
the surface must enclose a topological
defect. In the above example with $G=$U(1), 
the winding number along a closed curve
$C$ would be 
\begin{equation}
\label{equ:cont_wind}
N_W(C)=\frac{1}{2\pi}\oint_C d\vec{r}\cdot\vec{\nabla}\theta
\equiv \frac{1}{2\pi}\Delta\theta,
\end{equation}
where $\theta=\arg\phi$ is the phase angle of the order parameter
field $\phi$.

\subsection{Defects in gauge theories}
\noindent
In gauge field theories, the discussion of the vacuum manifold becomes
more subtle. A gauge transformation can be used to move the field 
anywhere in $\M$ at each point in space-time
independently
of all other space-time points. For instance, it is always possible to
carry out a gauge transformation into the {\it unitary gauge} in which the
direction of the order parameter is the same everywhere.
This does not by any means imply that there cannot be any topological
defects in gauge theories, because the transformation also changes the
gauge field. In fact, topological defects typically manifest 
themselves in the unitary gauge as non-physical 
gauge field singularities similar to the Dirac string
in the Dirac monopole configuration.

A simple example of a defect in a gauge theory is the Nielsen-Olesen
vortex\cite{Nielsen:1973cs} in the Abelian Higgs model. 
This model is a gauge invariant generalization of the global U(1)
theory in Eq.~(\ref{equ:phi4def}), and its Lagrangian is
\begin{equation}
\label{equ:AHdef}
{\cal L}=-\frac{1}{4}F_{\mu\nu}F^{\mu\nu}+D_\mu\phi^*D^\mu\phi-V(|\phi|).
\end{equation}
The Nielsen-Olesen solution can be
written in the form
\begin{equation}
\phi(r,\varphi,z)=ve^{iN_W\varphi}
f\left(\lambda/e^2,evr\right),\quad
A^i(r,\varphi,z)=-\frac{N_W}{er}\hat{\varphi}^ia(evr),
\end{equation}
where $v$ is the minimum of $V(|\phi|)$, $N_W$ is an integer and
$a(x)$ and $f(x)$ are functions, which must be determined
numerically. They vanish at $x\rightarrow 0$
and approach unity exponentially at $x\rightarrow\infty$.
If we calculate the total magnetic flux carried by the vortex, we find
\begin{equation}
\Phi=\lim_{R\rightarrow\infty}
\oint_{|\vec{r}|=R}d\vec{r}\cdot\vec{A}
=-\lim_{R\rightarrow\infty}2\pi R \frac{N_W}{eR}a(evR)
=-N_W\frac{2\pi}{e},
\end{equation}
which means that the magnetic flux is a multiple of the {\it flux
quantum} $\Phi_0=2\pi/e$.
The Abelian Higgs model is very similar to the Ginzburg-Landau theory
of super\-conductivity,\cite{Tilley} and in that context, the
Nielsen-Olesen vortices are known as Abri\-kosov flux tubes.

An important property of the Nielsen-Olesen solution is that the
energy density decreases exponentially at large $r$. 
In the global theory (\ref{equ:phi4def}), 
the energy density around a vortex has a power-law
decay $1/r^2$ because of the gradient term, but in the gauge theory it
is mostly cancelled by the gauge field contribution to the covariant
derivative $\vec{D}\phi$.
Therefore, the energy per unit length of a vortex, \ie, the tension,
is finite in the gauge theory but logarithmically divergent in the
global theory. This has the further implication that the interaction
of two gauge vortices is exponentially suppressed at long distances,
whereas global vortices have long-range interactions.

The interaction between two vortices 
is attractive if $\lambda/e^2<1/2$ and repulsive if 
$\lambda/e^2>1/2$.\cite{Jacobs:1979ch} These cases correspond to type-I
and type-II superconductors, respectively.\cite{Tilley}
Consequently, a solution with $|N_W|>1$ is only stable if $\lambda/e^2<1/2$.
In the special case $\lambda/e^2=1/2$, which is known as the Bogomolnyi
point,
one can find the exact form of the
functions $f(x)$ and $a(x)$, and calculate the vortex tension 
exactly.\cite{Bogomolny:1976de}

Another important defect solution is the 't~Hooft-Polyakov solution in
the Georgi-Glashow model.\cite{'tHooft:1974qc,Polyakov:1974ek} 
The theory consists of a Higgs field $\phi$
in the adjoint representation of the SU(2) gauge group. The Lagrangian
can be written as
\begin{equation}
{\cal L}=-\frac{1}{2}{\rm Tr}F_{\mu\nu}F^{\mu\nu}+{\rm Tr}[D_\mu,\phi]
[D^\mu,\phi]
-m^2{\rm Tr}\phi^2-\lambda{\rm Tr}\phi^4,
\end{equation}
where $D_\mu=\partial_\mu+igA_\mu$ and
$F_{\mu\nu}=(ig)^{-1}[D_\mu,D_\nu]$. Classically, the SU(2) symmetry
is broken into U(1) when $m^2$ is negative. The monopole solution has
the form\cite{'tHooft:1974qc,Polyakov:1974ek}
\begin{equation}
\phi=\frac{\vec{\sigma}\cdot\vec{r}}{gr^2}H(gvr),\qquad
\vec{A}=\frac{\vec{\sigma}\times\vec{r}}{gr^2}\left[1-K(gvr)\right],
\end{equation}
where $H(x)$ and $K(x)$ are functions which must be determined
numerically and which behave asymptotically as $H(x)\sim x$ and
$K(x)\rightarrow 0$.
Going to the unitary gauge $\phi\propto\sigma^3$, we can see that the
solution acts as a source for the magnetic field that corresponds to
the residual U(1) symmetry.\cite{'tHooft:1974qc}

Similar monopoles are also present if the group $G$ that is broken is more
complicated but has an SU(2) subgroup that is broken into U(1). 
In particular, practically any grand unified
theory (GUT) predicts the existence of magnetic 
monopoles.\cite{Preskill:1979zi} The
masses of these monopoles would be around the GUT scale,
which explains why they have not been observed in experiments. 
However, if the
temperature of the universe was initially above the GUT scale,
monopoles should have been formed in the phase transition in which the
GUT symmetry group broke down into SU(3)$\times$SU(2)$\times$U(1). 
This would have had disastrous effects
for the later evolution of the universe. 
Therefore, it is commonly believed that if monopoles were formed, they
were diluted away by a subsequent period of cosmological
inflation.\cite{Guth:1981zm} 

%\newpage
\section{Field theories at high temperatures}
\selabel{sect:finiteT}
\noindent
\subsection{Equilibrium}
\noindent
In Section~\ref{sect:Topo}, we discussed only classical solutions and 
smooth field
configurations, but in general, we shall be interested in phase
transitions that start from a thermal equilibrium state. 
The density operator in thermal equilibrium at
temperature $T=1/\beta$ is $\rho\propto\exp(-\beta H)$, where $H$ is the
Hamiltonian.
From this, we can in principle calculate the expectation
value of any observable
$X$,
\begin{equation}
\langle X\rangle={\rm Tr}\rho X.
\end{equation}
For practical calculations involving only static equilibrium quantities, 
it is often easier to use the imaginary
time formalism,
in which the observables are expressed as Euclidean
path integrals\cite{Kapusta} 
\begin{equation}
\langle X\rangle=Z^{-1}\int D\phi \exp\left(-\int_0^\beta d\tau\int d^3x
{\cal L}_{\rm Eucl}[\phi(\tau,\vec{x})]\right) X[\phi(0,\vec{x})]
\end{equation}
and the partition function $Z$ normalizes the result is such a way
that $\langle 1\rangle=1$.
The path integral has a similar form in gauge field theories as well,
although the temporal component $A_0$ of the gauge field needs a
special treatment.\cite{Gross:1981br}

In the high-temperature limit $\beta\rightarrow 0$, the compact
imaginary time dimension shrinks to a point. Therefore, at high
temperatures, the state of the system is well approximated by a
three-dimensional path integral\cite{Ginsparg:1980ef}
\begin{equation}
\label{equ:DR}
\langle X\rangle=Z^{-1}\int D\phi \exp\left(-\beta\int d^3x
{\cal L}_{\rm 3D}[\phi(\vec{x})]\right) X[\phi(\vec{x})].
\end{equation}
This approximation is called {\em dimensional reduction}, and it has
been used very successfully in the study of the electroweak phase
transition.\cite{Kajantie:1996dw,Kajantie:1997qd} 
The resulting path integral 
has exactly the same form as the corresponding expectation value
in classical statistical physics if we replace ${\cal L}_{\rm 3D}$ by
the classical Hamiltonian.

In thermal equilibrium, the state of the system is specified by the
expectation values of all the observables. In a finite volume, these
expectation values possess the same symmetries as the Lagrangian, and
therefore all observables that transform covariantly under symmetry
transformations vanish. However, in the thermodynamic limit, \ie,
in infinite volume, they may obtain non-zero expectation values, 
which signals that the symmetry is spontaneously broken. It depends on the
parameters whether this happens, and the partition function of the
system is non-analytic at the boundary between unbroken and broken
symmetries. In other words, the system has a phase transition.

In the classical zero-temperature case discussed in Section
\ref{sect:Topo}, 
it was easy to determine whether a symmetry was broken by
simply finding the global minimum of the potential. In the thermal
case (or in the zero-temperature quantum case), 
we can define an effective potential,\cite{Jackiw:1974cv} 
which has the same
property, by
\begin{equation}
\label{equ:Veffdef}
V_{\rm eff}(\varphi)=-\frac{T\ln\langle\delta\left(\overline{\phi}-
\varphi\right)
\rangle}{V},
\end{equation}
where $\overline\phi=V^{-1}\int d^3x\phi$.
This corresponds to the free energy density of a microcanonical 
system in which the 
volume average of the order parameter $\phi$ is fixed to the value
$\varphi$.
Because the free energy is minimized in thermal equilibrium, the
minimum of the potential shows the equilibrium state of the system.

In perturbation theory, the effective potential can be calculated
as a sum of one-particle irreducible (1PI) vacuum diagrams expanded around
the desired value of $\phi$.\cite{Jackiw:1974cv}
In the global quantum 
theory (\ref{equ:phi4def}), the one-loop effective potential is
\begin{eqnarray}
V_{\rm eff}(\varphi)&=&V(\varphi)+
\frac{T^2}{24}\left[\sum_i m_i^2(\varphi)+O(m(\varphi)^3/T)\right]
\nonumber\\
&=&V(\varphi)+\frac{1}{6}\lambda T^2|\varphi|^2+\frac{m^2T^2}{12}
+O(m(\varphi)^3T).
\end{eqnarray}
This contribution increases the effective mass of $\phi$ from $m^2$ to
$m^2+\lambda T^2/3$.
Therefore, even if the symmetry is broken at zero temperature, \ie,
$m^2<0$,
it will
be restored when the temperature becomes high enough $T>T_c\approx
(-3m^2/\lambda)^{1/2}$.

In gauge field theories, it is straightforward to show that the gauge
invariance is never spontaneously broken,\cite{Elitzur:1975im}
but this does not mean that there cannot be phase transitions.
In a fixed gauge, we can calculate $V_{\rm eff}(\varphi)$ perturbatively.
If the Higgs self-coupling $\lambda$ is much less than the gauge
coupling
$g^2$, this
computation is believed to be reliable and produces a non-analytic cubic
term $\propto|\varphi|^3$.\cite{Coleman:1973jx,HLM} 
This means that at a certain temperature $T_c$,
there are two degenerate vacua corresponding to zero and non-zero
values of $\varphi$.
Consequently, the expectation value of $\phi$ jumps discontinuously
from zero to a non-zero value when the temperature decreases, 
\ie, there is a first-order phase
transition. 

When $\lambda\gg g^2$, perturbation theory becomes unreliable, and the
line of first-order transitions ends in a second-order point. 
In some cases, the transition disappears completely at higher
$\lambda$, and the phases are smoothly connected.\cite{Fradkin:1979dv}
In others, a
second-order transition survives. This happens, for instance, in the
Abelian Higgs model; the transition at large $\lambda/e^2$ has
its most natural description in terms of a dual theory, whose
fundamental degrees of freedom are Nielsen-Olesen
vortices.\cite{Kleinert:1982dz,Kovner:1991pz} 

\subsection{Non-equilibrium}
\noindent
Formally, the time evolution of a quantum field theory 
is given by the operator equation of
motion
\begin{equation}
\label{equ:timeevol}
i\frac{\partial X}{\partial t}=[X,H].
\end{equation}
If $H$ is a constant in time, the system remains in thermal
equilibrium, and if $H$ depends on time, we obtain non-equilibrium
evolution. The problem is that $X$ is an operator, and therefore it is
extremely difficult to solve Eq.~(\ref{equ:timeevol}) in any
non-trivial field theory. We have to resort to approximations,
which are discussed in more detail in Section~\ref{sect:methods}.

%\newpage
\section{First-order phase transitions}
\selabel{sect:1storder}
\noindent
If the transition is of first order, the symmetric phase remains
metastable even slightly below the critical temperature. The effective
potential has two minima, and at the transition point, the one
corresponding to the broken phase becomes the global minimum. After
that, there is a finite probability per unit volume that the thermal
fluctuations throw the system over the potential barrier from the
metastable symmetric phase into the true, broken minimum. It is more
probable that this happens in only a small region of space. The
boundaries around this bubble of broken phase cost energy, but the
potential energy inside it is lower than in the symmetric
phase. If the bubble is large enough, the contribution from the
potential energy becomes larger than that from the bubble wall, and
the bubble starts to grow.

We can estimate the nucleation rate of these bubbles by
calculating the energy $E_c$ of a critical bubble, \ie, the lowest-energy
bubble that starts to grow instead of shrinking 
away.\cite{Langer:1967ax,Langer:1969bc,Coleman:1977py}
This gives the exponential suppression of the nucleation rate,
$\Gamma_{\rm nucl}\propto\exp(-\beta E_c)$, and
therefore a reasonable order-of-magnitude estimate as long as $\beta
E_c\gg 1$. Calculating the
prefactor requires significantly more work.\cite{Moore:2001jw}

In a large system, many bubbles are nucleated at different points in
space, and when the bubbles grow, they eventually coalesce and fill
the whole space, thus completing the transition.
The bubble walls are typically moving relatively
slowly\cite{Kurki-Suonio:1996rk}   and are
preceded by shock waves, which heat up the
system.\cite{Ignatius:1994qn}
This suppresses the further
nucleation of bubbles nearby.

\subsection{Global theories}
\noindent
When a symmetry breaks, no point on the vacuum manifold $\M$ is
preferred over any other. In a first-order transition, it is therefore
a very reasonable assumption that the direction of symmetry breaking 
is uncorrelated between different bubbles.
Inside a single bubble, on the other hand, 
it is energetically preferable for the order
parameter field $\phi$ to be constant.
If the vacuum manifold $\M$ is not connected, \ie, $\pi_0(\M)\ne 1$, 
$\phi$ may belong to
different connected parts inside two neighbouring 
bubbles, and when the bubbles
meet, a domain wall is formed between them.\cite{Zeldovich:1974uw}

If $\M$ is connected but not simply connected, \ie, $\pi_1(\M)\ne 1$, then
$\phi$ can always relax to a constant after a collision of two
bubbles. However, if a third bubble hits the first two before this has
happened,
a vortex
can be formed. Imagine, for instance, that $G$=U(1) and
inside the three bubbles the phase angle $\theta$ of the order
parameter $\phi$ has the values $\theta_1$, $\theta_2$ and
$\theta_3$. It is common to assume  
the {\em geodesic rule},\cite{Kibble:1976sj,Vachaspati:1984dz} \ie, that
when the phase angle interpolates between the values inside two
bubbles, it always uses the shortest path in the vacuum manifold.
It may then happen that the path that connects $\theta_1$ to
$\theta_2$ to $\theta_3$ and finally to $\theta_1$ winds around the
vacuum manifold, in which case the order parameter cannot relax to a
constant everywhere and a vortex is formed.
Similarly, we can see that in theories with  $\pi_2(\M)\ne 0$,
collisions of four bubbles may lead to formation of monopoles. 

In this simple picture, the probability of forming a topological defect in a
bubble collision depends essentially only on the geometry of the
collision and on the geometry of the symmetry group. 
Therefore, we can conclude that the typical
distance between the vortices is proportional to the typical
separation between the bubble nucleation sites $d_{\rm nucl}$. 
We can characterize the density of topological defects whose
codimensionality is $D_{\rm co}$ by the number density $n$ of points where a
defect crosses a subspace of dimensionality $D_{\rm co}$.
According to the above discussion, it behaves as
\begin{equation}
\label{equ:kibbledens}
n\approx d_{\rm nucl}^{-D_{\rm co}}.
\end{equation}
In practice,
$D_{\rm co}$ is 1 for domain walls, 2 for vortices and 3 for monopoles.
For instance, the number density of vortices per unit cross-sectional
area is predicted to be $n\approx d_{\rm nucl}^{-2}$.

The probabilities of
forming defects 
in symmetric collisions of bubbles
have been
calculated by Prokopec.\cite{Prokopec:1991ab}
Although bubble nucleation is
essentially a random event and therefore the bubble
nucleation events are more or
less evenly distributed in space and time, it was customary in early
studies to assume that all bubbles were nucleated simultaneously in a
regular lattice.\cite{Vachaspati:1984dz} 
Obviously, the form of the
lattice then affects the results.\cite{Scherrer:1986sw} 
One attempt to avoid this
problem was the ``lattice-free'' approach\cite{Borrill:1995gu,deLaix:1999xz}
in which the bubble locations are random.

For a more accurate treatment, the field dynamics
must be taken into account. If the bubbles do not
collide simultaneously, the probability of forming a defect decreases,
because the order parameter may already have equilibrated before
the third bubble hits the first two.~\cite{Melfo:1995cv}
Furthermore, the growth of the bubbles is typically much slower than
the speed of light because of the friction
caused by the hot plasma that fills the space at high 
temperatures.\cite{Kurki-Suonio:1996rk}
This allows more time for
the phase equilibration between the first two bubbles to happen and therefore
decreases the probability of forming a defect 
even further,\cite{Ferrera:1996ef} 
although it also means that a larger fraction of space is in the
symmetric phase at any given time and therefore the total number of
bubble nucleation events is larger.

\subsection{Gauge field theories}
\selabel{sect:1stgauge}
\noindent
In gauge field theories, the above picture cannot be used directly,
because the direction of the order parameter is not gauge
invariant. In particular, we can always choose a gauge in which
$\phi$ is equal in the two colliding bubbles.
That led Rudaz and Srivastava\cite{Rudaz:1993wy} 
to question the applicability of the
geodesic rule in gauge theories.

The first detailed studies of the field dynamics in the 
Abelian Higgs model\cite{Hindmarsh:1994av,Kibble:1995aa} 
showed that the geodesic rule still holds 
if the phase transition starts  from a classical vacuum
state. 
In order to treat the phase angles in a gauge-invariant way,
Kibble and Vilenkin\cite{Kibble:1995aa}
introduced a
gauge-invariant phase difference
\begin{equation}
\label{equ:local_delta}
\Delta_{\rm gi}\theta=\int d\vec{r}\cdot \vec{D}\theta,
\end{equation}
where $\vec{D}\theta=\vec{\nabla}\theta+e\vec{A}$. 
This cannot be used as a direct substitute for the phase
angle difference defined in
Eq.~(\ref{equ:cont_wind}), 
because it
depends on the path along which the difference is calculated and
the phase difference around a closed curve is not quantized.
Furthermore, $\Delta_{\rm gi}\theta$ around a classical vortex
configuration is zero, not $2\pi$ as in the global case, and therefore
the gauge-invariant 
phase difference by itself cannot be used to locate vortices.

\begin{figure}
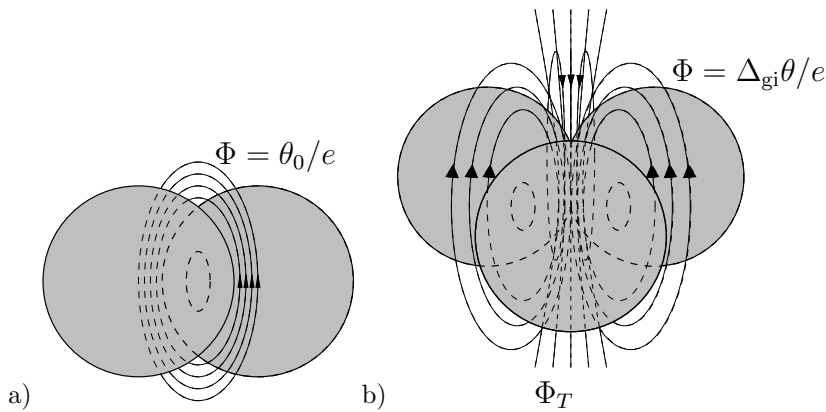

\center
a) \input 2bubbles_2.pstex_t
b) \input 3bubbles_2.pstex_t
\flushleft
\caption{
\label{fig:coll}
Bubble collisions in the Abelian Higgs model.
a) In the classical vacuum $\vec{A}=0$, a 
collision of two bubbles with a phase difference $\theta_0$ 
forms a ring of magnetic field.
b) In a collision of three bubbles, the flux rings and the initial
flux $\Phi_T$ between the bubbles contribute to the total flux which
turns into vortices.
}
\end{figure}

Let us first consider a 
collision of two bubbles in a 
classical vacuum, where the magnetic field vanishes, \ie,
$\vec{B}=\vec{\nabla}\times\vec{A}=0$.
In this case,
we can choose the gauge $\vec{A}=0$, which means that
the gauge invariant phase difference is simply the
ordinary phase difference.
If we denote
it by $\theta_0$, a circular magnetic flux
ring with flux $\Phi=\int d\vec{S}\cdot\vec{B}=\theta_0/e$ 
is formed around the point of their
collision, because the phase difference induces an electric 
current from one
bubble to another\cite{Rudaz:1993wy,Kibble:1995aa} 
(see Fig.~\ref{fig:coll}a).
In a simultaneous collision of three bubbles, all of the
three collisions are at first totally independent and in each of them
a flux ring like this is formed. Before the collision is complete, a
hole remains at the middle of the collision, and the total flux
$\Phi_{\rm tot}$
through this hole is the sum of the gauge invariant phase
differences divided by~$e$, \ie, $\Phi_{\rm tot}=\Delta_{\rm
gi}\theta/e$. When the hole shrinks, this flux is
squeezed into a Nielsen-Olesen vortex of winding $N_W=e\Phi_{\rm
tot}/2\pi=\Delta_{\rm gi}\theta/2\pi$. 
Because in the $\vec{A}=0$ gauge, $\Delta\theta=\Delta_{\rm
gi}\theta$, this is equal to the result (\ref{equ:cont_wind})
predicted by a naive application of the
geodesic rule.

The above discussion assumes that the fields equilibrate immediately
when the bubbles collide, which is not always a good approximation. It
was shown by Copeland and Saffin\cite{Copeland:1996jz} that in certain
cases the geodesic rule can be violated by violent non-equilibrium
effects after the collision. This happens, for instance, if the
gauge-invariant phase difference $\theta_0$ is large and the
bubbles are far apart, even if the transition starts from classical vacuum.

However, in practice the transition starts from a thermal state, and 
Kibble and Vilenkin\cite{Kibble:1995aa} pointed out that then the
fluctuations of the magnetic field are always
present. These fluctuations change the above 
picture, leading to a violation of the
geodesic rule.
More precisely, if the area between the bubbles is $A$,
the typical magnetic flux will be $\Phi_T\sim \sqrt{AT}$. 
Suppose for simplicity that the flux is peaked around the centre of the
collision so that it does not affect the individual two-bubble
collisions. Then, the three flux rings are again formed, and they
merge with the initial flux $\Phi_T$, which is trapped between the
bubbles (see Fig.~\ref{fig:coll}b).
This leads to the formation of a vortex with
winding
\begin{equation}
\label{equ:NWdefgauge}
N_W=(\Delta_{\rm gi}\theta-e\Phi_T)/2\pi=N_W^0-e\Phi_T/2\pi,
\end{equation} 
where $N_W^0$ is the winding
number predicted by $\Delta_{\rm gi}\theta$ alone.
Note that $N_W^0$ is typically not an integer, but $N_W$ always is.
If the bubbles are far apart, $\Phi_T$ can be very
large, and therefore vortices with very high windings can form.

The generalization of the above picture to non-Abelian gauge field
theories is still an open question. It is not at all obvious what the
analogue of the magnetic flux $\Phi_T$
in the case of, say, 't~Hooft-Polyakov monopoles is. Furthermore,
because of confinement, a non-Abelian gauge field theory cannot be
approximated by a free field theory. In this paper, we shall
therefore only consider Abelian gauge field theories.

%\newpage
\section{Continuous phase transitions}
\selabel{sect:2ndorder}
\noindent
The dynamics of first-order phase transitions are simplified by the
hierarchy between the bubble nucleation rate and other time
scales. Furthermore, because of the metastability we were allowed
to assume that
the temperature is initially below $T_c$ and that it remains constant
during the phase transition. In continuous transitions, the dynamics
of the transition depends on how, and especially how rapidly, the
phase transition line is crossed.

A continuous phase transition is usually of second order, which
means that the second derivative of the free energy is discontinuous
and at least one correlation length diverges.
Most commonly, these transitions are associated with a
spontaneous breakdown of a symmetry, but opposite examples are also
known.
For
instance, 
in the
two-dimensional XY model,\cite{Kosterlitz:1973xp} the classical
analysis {\it \`a la} Section~\ref{sect:SSB} 
would suggest that the symmetry gets broken, 
but a proper treatment shows that this is not the case, because a
continuous symmetry cannot be broken in a two-dimensional 
theory.\cite{MerminWagner,Coleman:1973ci}
A similar theorem applies to 
gauge field theories in any number of dimensions,\cite{Elitzur:1975im} 
and therefore depending on the model, the
phase transitions can range from first-order
transitions\cite{Coleman:1973jx}
 to smooth
crossovers.\cite{Fradkin:1979dv}

In the setup we shall consider, the system is initially in thermal
equilibrium in the symmetric phase. The system is forced to
undergo a phase transition by changing one of the parameters
continuously through the phase transition line at a rate
characterized by the {\em quench timescale} $\tau_Q$. 
In different cases, the parameter that is changed can be, for instance,  
a coupling
constant, the mass parameter, the scale factor of the universe or the
pressure.
Although the precise way the parameters are changed
certainly affects the detailed dynamics, the qualitative picture and
the order-of-magnitude estimates are believed to be unaffected.

The defects are counted after the transition when they
have become well-defined objects. If the interaction
between the defects has a short range, this can be done very reliably
by letting the system equilibrate locally to the low temperature so
that thermal fluctuations are negligible and counting the defects only
after that. When the interaction has a long range, the number of
defects after the transition has a certain ambiguity depending on how
they are counted, but because we are only interested in orders of
magnitude, we shall ignore these problems.

In most theoretical studies, it is assumed that the phase transition
is homogeneous, but in a real experiment any method of cooling or changing some
external parameters in some other way always leads to
inhomogeneities, which may be important for defect 
formation.\cite{KibbleInhomog,Grisha2000} 
We shall not discuss these complications here.

\subsection{Global theories}
\noindent
The overall picture of defect formation in second-order transitions in
global theories is in many ways similar to the first-order transitions
discussed in Section~\ref{sect:1storder}. After the transition,
the order parameter
can only be
correlated at distances less than some finite correlation length
$\hat{\xi}$
and therefore we can replace the bubbles in the analysis with correlated
domains of radius $\hat{\xi}$.\cite{Zeldovich:1974uw,Kibble:1976sj}
When we replace $d_{\rm nucl}$ by $\hat{\xi}$ in
Eq.~(\ref{equ:kibbledens}), we find
\begin{equation}
n\approx\hat{\xi}^{-D_{\rm co}}.
\end{equation}
In cosmology, we can state the ultimate upper limit for $\hat{\xi}$
as the age of the universe,\cite{Zeldovich:1974uw} 
but in practice $\hat{\xi}$ is
always much shorter than that. 

The early suggestion\cite{Kibble:1976sj}  was to identify
$\hat{\xi}$ with the correlation length at the Ginz\-burg temperature
$T_G$ defined by
\begin{equation}
\label{equ:Tginzburg}
\frac{\Delta V(T_G)}{T_G}\xi(T_G)^3=1.
\end{equation}
Below $T_G$, the thermal fluctuations become incapable of
restoring the symmetry in patches larger than the correlation
length. This would mean that the final number
density of defects is independent of the quench timescale $\tau_Q$. 
However, symmetry restoration in a patch of radius $\hat{\xi}$ can
only form vortex loops of radius less than $\hat{\xi}$ or
monopole-antimonopole pairs with separation less than $\hat{\xi}$, and
these are unstable configurations which quickly annihilate
when the symmetry gets broken again. The defects that survive after
the transition must have therefore been formed at an earlier
stage. Nevertheless, the Ginzburg temperature still has the significance
that individual defects can only be identified reliably at $T<T_G$.

The attempt to use the Ginzburg temperature to estimate the defect
density also
ignores the fact that reordering the order parameter at long distances
takes time.\cite{Kibble:1980mv}
Therefore, even if it were energetically possible, it may
not have have time to happen in a rapid transition. Instead, the
relevant length scale is actually determined as a compromise between
the tendency of ordering the fields at long distances and the finite
time available for that process. This argument was used by Zurek to
derive a more realistic estimate for the defect 
density.\cite{Zurek:1985qw,Zurek:1993ek}

A characteristic property of second-order phase transitions is that
the correlation length of the order parameter diverges at the
transition point with some critical exponent $\nu$, \ie,
$\xi(T)\approx\xi_0|\epsilon|^{-\nu}$, where $\epsilon(T)=(T_c-T)/T_c$
and
the exponent $\nu$ depends only on the universality class of the
model. However, in practice, $\xi$ cannot grow arbitrarily fast,
because at least it is constrained by causality to change slower than
the speed of light.\cite{Kibble:1980mv} 
If the transition takes place at a finite rate,
say, $\epsilon=t/\tau_Q$, then eventually a point is reached at which
the true correlation length cannot keep up with the equilibrium
correlation length $\xi(T)$ at the same temperature. After that, the
dynamics of the system cease to be adiabatic.

In reality, other effects make the maximum growth rate of the
correlation length much less than the speed of
light. Zurek\cite{Zurek:1985qw} 
assumed
that the relevant dynamics of the system can be characterized by a
relaxation timescale $\tau(T)$, which also diverges at the
transition point in equilibrium, but typically with a different
critical exponent, $\tau\approx\tau_0|\epsilon|^{-\mu}$.

Zurek's estimate is based on the idea that when the transition is
approached from the symmetric phase, the dynamics of the system
eventually become too slow to stay adiabatic and the maximum
correlation length $\hat{\xi}$ is roughly the correlation
length at that time. More precisely, he used the condition
$\tau(\hat{t})=|\hat{t}|$ to signal the breakdown of adiabaticity,
which implies
\begin{equation}
\hat{t}\approx-(\tau_0\tau_Q^\mu)^{1/(1+\mu)}.
\end{equation}
The maximum correlation length is then
\begin{equation}
\label{equ:zurekpred0}
\hat{\xi}\approx\xi_0\left(\frac{\tau_Q}{\tau_0}\right)^{\nu/(1+\mu)}.
\end{equation}

The critical exponents $\mu$ and $\nu$ depend on the system, and
at intermediate length scales, they may be different from their true,
asymptotic values.
Laguna and
Zurek\cite{Laguna:1998cf}
have
discussed two special cases, the overdamped case with $\mu=1$ and the
underdamped case with $\mu=2$.
In mean-field theory $\nu=1/2$, implying
\begin{equation}
\label{equ:zurekpred}
\hat{\xi}_{\rm MF}(\tau_Q)\approx\left\{
\begin{array}{ll}
\xi_0\left(\tau_Q/\tau_0\right)^{1/4}, & {\rm (overdamped)}\cr
\xi_0\left(\tau_Q/\tau_0\right)^{1/3}. & {\rm (underdamped)}\cr
\end{array}
\right. 
\end{equation}
In any case, causality
constrains $\mu\ge\nu$.

Finally, let us clarify the meaning of the correlation length by
discussing it from another viewpoint. If we denote the dimensionality of
the order parameter by $N$, there
are $N$ orthogonal elementary excitations. In the symmetric phase,
they are all degenerate, \ie, have the same correlation
length. Therefore it makes sense to talk about ``the correlation
length'' $\xi$. However, in equilibrium in the broken phase, only one of these
correlation lengths is finite, namely the one that corresponds to
the modulus of the order parameter. 
The direction of the order parameter becomes a Goldstone mode
with an infinite correlation length. 

After a rapid phase transition, both correlation lengths are initially
roughly equal, $\xi_{\rm mod}$, which corresponds to the modulus of
the order parameter, and $\xi_{\rm GS}$, which corresponds to the
Goldstone modes. However, $\xi_{\rm mod}$ starts to decrease rapidly,
and because  it determines the size of the defect core, the defects become
well-defined objects. On the other hand, $\xi_{\rm GS}$ characterizes 
the separation between the defects, since it is easy to see that
defects cause disorder to the order parameter field at distances larger
than their separation, but at shorter distances the direction of the 
order parameter is correlated. When the defect configuration evolves,
defects annihilate, and $\xi_{\rm GS}$ approaches its infinite
equilibrium value, albeit very slowly. This is another way of seeing
why the maximum correlation length should determine the density of
defects.

In the following we shall call the above scenario the {\it Kibble-Zurek
mechanism}. It is believed to be generally valid in phase transitions
in which a global symmetry is broken. During the recent years,
it has been tested extensively in numerical simulations and condensed
matter experiments, which we shall discuss in more detail in
Sections~\ref{sect:methods} and \ref{sect:expt}, respectively.

\subsection{Gauge field theories}
\selabel{sect:HRmech}
\noindent
As in the case of first-order phase transitions, the gauge fields were
long thought to be irrelevant for defect 
formation.\cite{Hindmarsh:1995re} This view
was even supported by early numerical simulations,\cite{Yates:1998kx} 
which showed a good
agreement with the Kibble-Zurek mechanism. It is now understood that
this was because the initial temperature was very low in 
them.\cite{Hindmarsh:2000kd,StephensNew}

Just like in the case of first-order transitions, thermal
fluctuations of the magnetic field are present
at any non-zero initial temperature, and they have an important effect
on the dynamics.
As discussed in Section~\ref{sect:1stgauge}, the magnetic flux that
originates in thermal fluctuations can get trapped between the bubbles
in three-bubble collisions in a first-order phase transition, 
leading to the formation of vortices with higher
windings.\cite{Kibble:1995aa}
In a similar way, Zurek\cite{Zurek:1996sj} argued that when a
superconductor loop is rapidly quenched into the superconducting
phase, magnetic flux can get trapped inside even when the transition
is continuous. In that case, it
cannot be interpreted as a vortex, though, because its core is outside
the system.

The dynamics are more complicated in the ``bulk'' case, in which the 
system fills the whole space. 
It was shown by Hindmarsh and
Rajantie\cite{Hindmarsh:2000kd}
that even in a continuous phase transition, the long-wavelength
thermal 
fluctuations of the magnetic field freeze out
in the phase transition and form vortices. 
More recently, similar conclusions were also reached by Stephens {\it et
al.},\cite{StephensNew} who studied instantaneous quenches.
We
shall call this way of
forming vortices the {\it flux trapping} mechanism.

The starting point of the flux trapping mechanism is that 
in the Coulomb (normal)
phase, there are fluctuations of the magnetic field with arbitrarily
long wavelengths. The simplest approximation is to treat the Coulomb
phase as a vacuum, in which case these thermal fluctuations are
simply blackbody radiation. Provided that the coupling to the other
fields is weak, the fluctuations are Gaussian and uncorrelated, and 
the width of the Gaussian distribution is
proportional to the temperature and independent of the momentum. 

The local magnetic field $\vec{B}$ is given by $\vec{\nabla}\times\vec{A}$.
If we define its two-point correlation function $G(k)$ as
\begin{equation}
\left\langle B_i(\vec{k})B_j\vec({k}')\right\rangle=
\left(\delta_{ij}-\frac{k_ik_j}{k^2}\right)
(2\pi)^3\delta\left(\vec{k}+\vec{k}'\right)
G(k),
\end{equation}
we have in the free-field approximation
\begin{equation}
\label{equ:symmcorr}
G(k)=T.
\end{equation}
Of course, neither the high-temperature phase of the Abelian Higgs
model nor the the normal phase of a superconductor is exactly a
vacuum, and therefore Eq.~(\ref{equ:symmcorr}) obtains corrections,
which suppress the fluctuations. Near the transition point, critical
fluctuations also change the correlator, and the magnetic field
obtains a non-trivial anomalous dimension. Therefore we can generally
assume that in the limit $k\rightarrow 0$,
\begin{equation}
\label{equ:symmcorrnu}
G(k)\propto k^\eta,
\end{equation}
with some anomalous dimension $0\le\eta<2$.

In the Higgs (superconducting) phase, long-range fluctuations are
suppressed, and the equilibrium two-point function becomes
\begin{equation}
\label{equ:brokencorr}
G(k)\approx \frac{Tk^2}{k^2+m_\gamma^2},
\end{equation}
where $m_\gamma$ is the inverse correlation length of the magnetic
field,
\begin{equation}
m_\gamma\approx ev\approx\sqrt{\frac{e^2}{\lambda}(-m^2)}.
\end{equation}

When the system goes through the phase transition,
each Fourier mode of the magnetic field must change its amplitude from
Eq.~(\ref{equ:symmcorr})  into
Eq.~(\ref{equ:brokencorr}) in order to remain in equilibrium. 
However, it has only a finite amount of time available for this, and
if it reacts too slowly, it falls out of equilibrium. 
The crucial point is that the modes with long wavelengths react slower
and this is therefore more likely to happen to them.
We denote by $\tau(k)$ the
time scale within which a Fourier mode with wavenumber $k$ can 
change, and in the vacuum approximation it is simply the
inverse frequency, $\tau(k)\approx k^{-1}$. More realistically, the time
scale may be determined by the conductivity of the material, or by
plasma effects, but in any case, $\tau(k)\rightarrow\infty$ as
$k\rightarrow 0$, which means that there is always some critical
wavelength $\hat{\xi}$ so that modes with longer wavelengths (\ie,
$k\lsim \hat{\xi}^{-1}$) cannot adjust but
fall out of equilibrium.

What this means is that after the phase transition, the
distribution of the magnetic field at distances longer than
$\hat{\xi}$ is the same as it was before the transition. In
particular, we can calculate the typical (rms) 
magnetic flux through a loop
of radius $\hat{\xi}$ using Eq.~(\ref{equ:symmcorrnu}). The result
depends on the dimensionality $D$ of the space, and assuming $\eta=0$,
we
find\cite{Hindmarsh:2000kd}
\begin{equation}
\label{equ:fluctPhi}
\Phi_{\hat{\xi}}\approx T^{1/2}\hat{\xi}^{2-D/2}.
\end{equation}
As the modes with wavelength less than $\hat{\xi}$ were able to
equilibrate, the field configuration inside this loop has relaxed into
its minimum energy configuration, which is a cluster of
\begin{equation}
\label{equ:cluster}
N_{\hat{\xi}}\approx\Phi_{\hat{\xi}}/\Phi_0
\approx\frac{e}{2\pi}T^{1/2}\hat{\xi}^{2-D/2}
\end{equation}
vortices, each having the same sign.
Dividing this by the area of the loop, we find the number density of
vortices formed in the transition,
\begin{equation}
\label{equ:HRdens}
n\approx \frac{N_{\hat{\xi}}}{\hat{\xi}^2}
\approx \frac{e}{2\pi}T^{1/2}\hat{\xi}^{-D/2}.
\end{equation}
It is important to note that the dependence on $\hat{\xi}$ is
different from the prediction (\ref{equ:kibbledens}) of the Kibble
mechanism, and that the density increases with increasing temperature
and increasing gauge coupling constant. In particular,
Eq.~(\ref{equ:HRdens}) vanishes in the global limit $e\rightarrow 0$, and only 
the Kibble-Zurek mechanism remains.

It has been argued by Stephens {\it et al.}\cite{StephensNew}
that if the phase transition is very rapid, the picture becomes
more complicated. In the Higgs phase, the magnetic field can only
penetrate the system if it exceeds the critical field $B_{\rm c1}$.
Near the transition, $B_{\rm c1}$ vanishes, but if the quench is
instantaneous, we can assume that it jumps discontinuously to a
non-zero value. This would mean that no
vortices are formed in 
regions where the flux density is lower than $B_{\rm c1}$.

\begin{figure}
\center
\epsfig{file=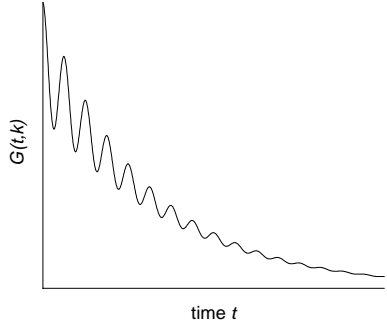,width=5cm}
\flushleft

\caption{
\label{fig:Landau}
Typical form of the real-time correlator $G(t,k)\propto\langle
\vec{B}(0,\vec{k})\vec{B}(t,-\vec{k})\rangle$ of the magnetic field.
It is a linear combination of an exponentially
decreasing Landau damping contribution $\exp(-\gamma_L t)$ and an
oscillatory plasmon contribution $\exp(-\gamma_p t)\cos(\omega_p t)$.
At long wavelengths, $\gamma_L\ll \omega_p,\gamma_p$,
\protect\cite{Hindmarsh:2001vp} and therefore $\gamma_L$
determines the time $\tau\sim\gamma_L^{-1}$ when $G(t,k)$
vanishes. This is the time needed for a perturbation of the magnetic
field to equilibrate.
}
\end{figure}

In order to use Eq.~(\ref{equ:HRdens}) to predict 
the number density of defects, we
must be able to calculate $\hat{\xi}$. 
Hindmarsh and Rajantie\cite{Hindmarsh:2000kd}
argued that this can be found by using an adiabaticity condition, 
which balances the rate at which the
equilibrium two-point function $G(k)$ of a given Fourier mode
must change in order to remain
in equilibrium with the dynamical time scale of the same mode,
\begin{equation}
\label{equ:adiab}
\left|\frac{d\ln G(\hat{k})}{dt}\right|=\tau^{-1}(\hat{k}).
\end{equation}
Furthermore, 
they argued that in the relativistic case, $\tau(k)$ is determined by
the Landau damping rate\cite{Hindmarsh:2001vp} $\gamma_L$ 
[see Fig.~\ref{fig:Landau}]
\begin{equation}
\label{equ:pertlandau}
\tau^{-1}(k)=\gamma_L(k)\approx\frac{4k^3}{\pi m_D^2}.
\end{equation}
Assuming that $G(k)$ is given by Eq.~(\ref{equ:brokencorr}), where
$m^2=-\delta m^2 t/\tau_Q$, we find that Eq.~(\ref{equ:adiab}) becomes
\begin{equation}
\frac{e^2}{\lambda}\frac{\delta m^2}{\tau_Q}\frac{1}{\hat{k}^2}\approx
\frac{4\hat{k}^3}{\pi m_D^2},
\end{equation}
from which we obtain
\begin{equation}
\label{equ:HRhatxi}
\hat{\xi}\propto\hat{k}^{-1}\propto\tau_Q^{1/5}.
\end{equation}
Using Eq.~(\ref{equ:HRdens}), we can then see
that the number density of vortices per unit cross-sectional area
behaves in three spatial dimensions as
\begin{equation}
n\propto\tau_Q^{-0.3}.
\end{equation}
However, this result is not valid at very 
large $\tau_Q$.\cite{Hindmarsh:2001vp} When $k\ll e^2T$, 
non-perturbative effects change the form of the equilibrium
correlator $G(k)$, leading to a non-trivial anomalous dimension
$\eta$. Furthermore, at low $k$, it is conductivity $\sigma$ rather
than Landau
damping that determines the rate of exponential damping $\gamma_L$ 
and this changes its behaviour to
$\gamma_L\approx k^2/\sigma$.

It is obvious, too, that while the equilibrium equal-time correlator
$G(k)$ is believed to be universal, the dynamics and therefore the
relaxation time scale $\tau(k)$ are sensitive to the details of the
system. This means that the power law may be different for, say,
superconductors, although the mechanism itself is believed to be valid.
Finally, when applying Eq.~(\ref{equ:HRdens}) to a transition in a
superconducting film, we have to be careful, because although the film
itself is two-dimensional, the magnetic field extends to the third
dimension as well.\cite{Rajantie:2001na}

\subsection{Spatial distribution of topological defects}
\selabel{sect:spatdist}
\noindent
Because neither the Kibble-Zurek nor the flux trapping mechanism 
can give clear-cut predictions for
the number density of topological defects formed in a phase
transition unless we make certain, fairly strong assumptions about the
dynamics, the number density itself is not a very good quantity for
testing the mechanisms. Instead, we should look for signals that are
truly qualitatively different in the two mechanisms, such as the
spatial distribution of vortices after the phase transition.
This has been emphasized by Digal {\it et al.}\cite{Digal:1998ak}
who pointed out that the Kibble mechanism leads to strong negative
vortex-vortex correlations.

A simple way of seeing that the two mechanisms predict very different
spatial distributions is to look at the quantity
$N_C(r)$,\cite{Hindmarsh:2000kd} defined as the average winding
number around a circle of radius $r$ centred at the positive
vortex. Formally, we may write it as the expectation value
\begin{equation}
\label{equ:NCr}
N_C(r)=\left\langle n(\vec{x})\frac{1}{2\pi}
\oint_{|\vec{x}'-\vec{x}|=r}
d\vec{x}'\cdot\vec{\nabla}\theta(\vec{x}')
\right\rangle
=\left\langle n(\vec{x})
\int_{|\vec{x}'-\vec{x}|\le r}
d^2x' n(\vec{x}')
\right\rangle,
\end{equation}
where $n(\vec{x})$ is the winding number density at point $\vec{x}$.
From the latter form we obtain
\begin{equation}
\label{equ:dNCr}
\frac{dN_C(r)}{dr}=
2\pi r
\left\langle n(0)
n(r)
\right\rangle.
\end{equation}

\begin{figure}
\center
a)
\epsfig{file=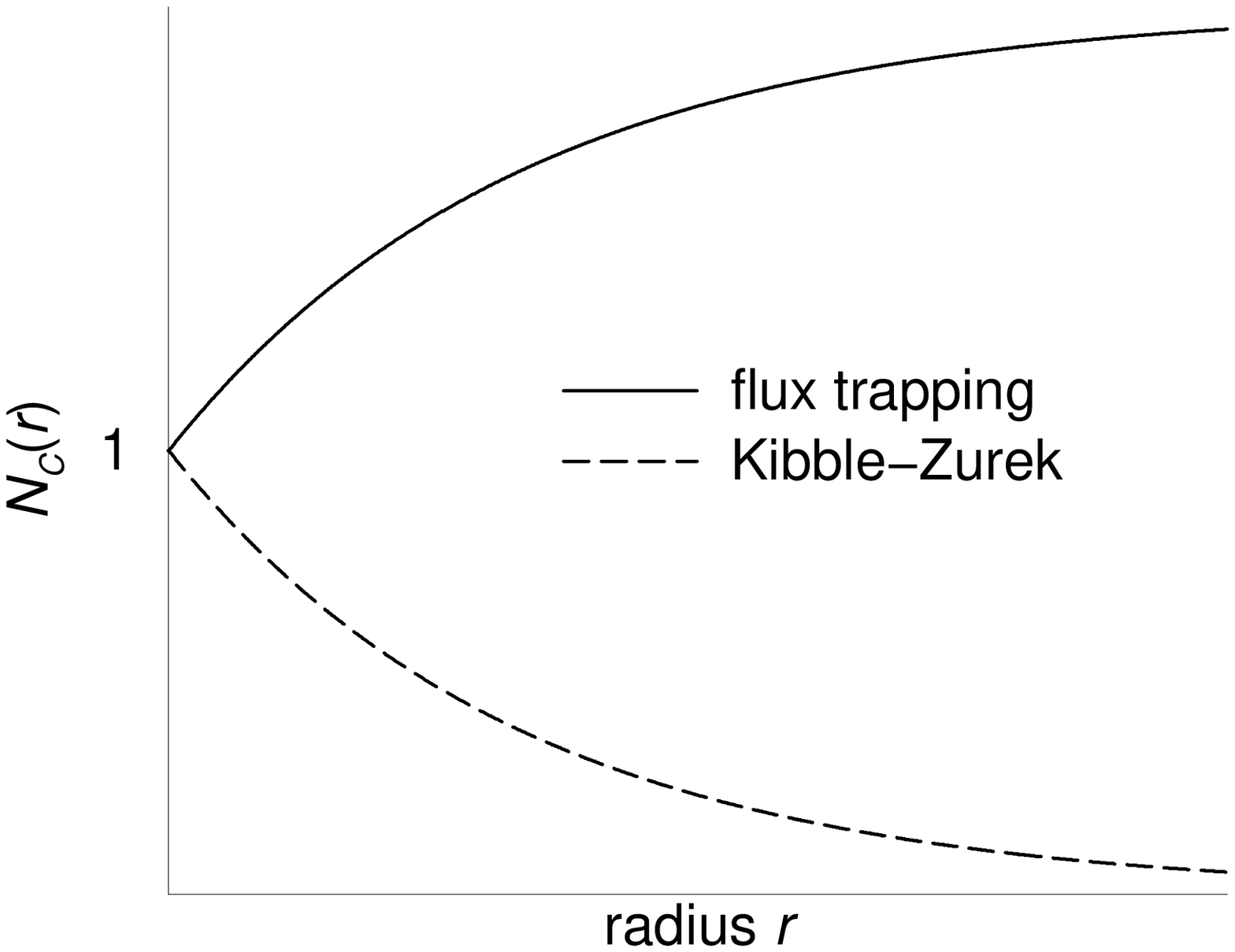,width=5cm}
b)
\epsfig{file=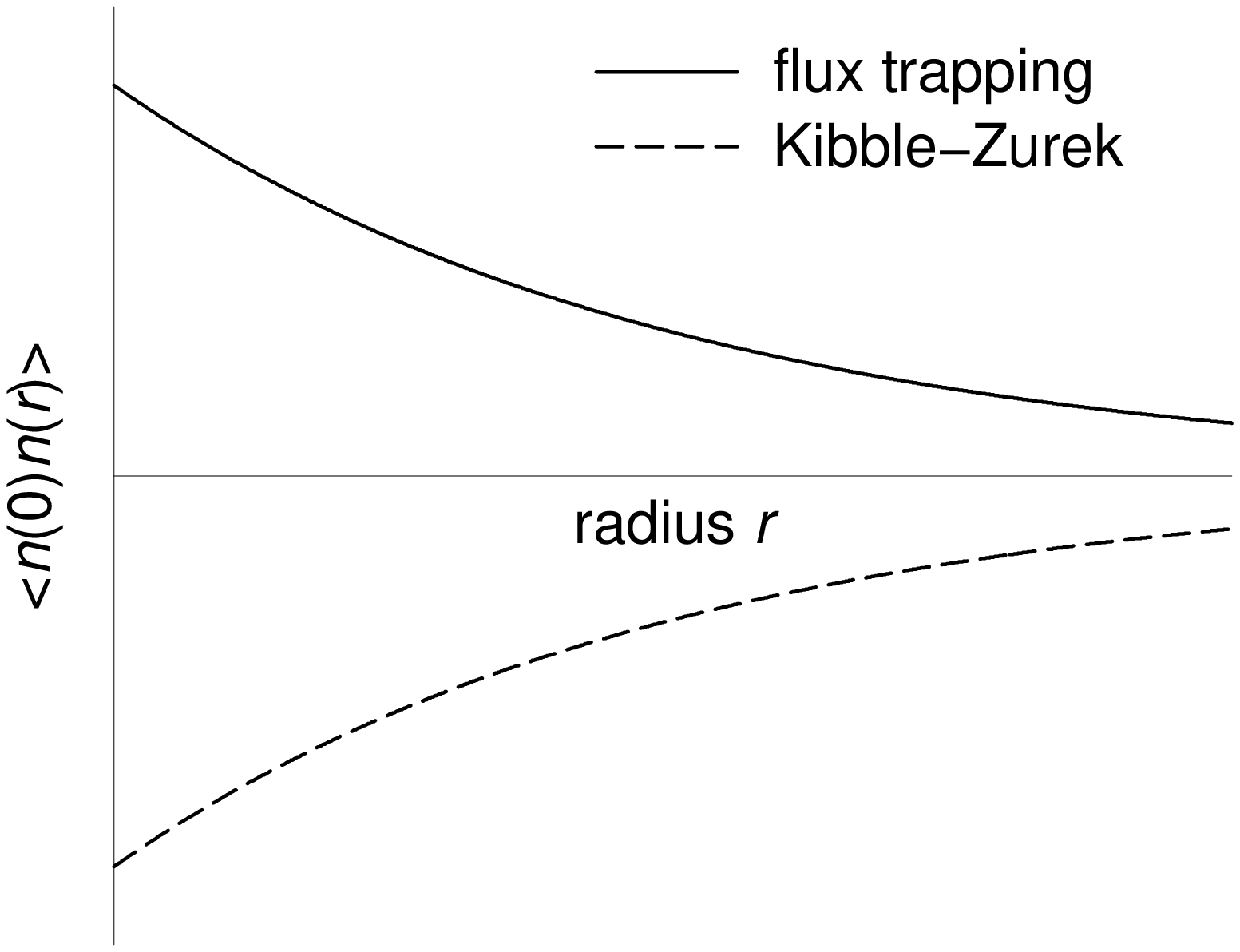,width=5cm}
\flushleft
\caption{
\label{fig:ncorr}
a) The behaviour of $N_C(r)$ defined in Eq.~(\ref{equ:NCr}) as predicted by
the flux trapping and Kibble-Zurek mechanisms.
b) The form of the vortex-vortex correlation function $\langle n(0)n(r)
\rangle$
as predicted by the flux trapping and Kibble-Zurek mechanisms.
}
\end{figure}

Let us first consider the Kibble-Zurek mechanism. 
The typical separation of vortices is given by $\hat{\xi}$, whereby
we do not encounter any
vortices of either sign at short distances, $r\ll\hat{\xi}$,
as we increase $r$.
However, when $r\gg\hat{\xi}$, this changes, because we can write
Eq.~(\ref{equ:NCr}) in the form
\begin{equation}
N_C(r)=\frac{1}{2\pi}
\oint_{|\vec{x}'-\vec{x}|=r}
d\vec{x}'\cdot
\left\langle n(\vec{x})\vec{\nabla}\theta(\vec{x}')
\right\rangle,
\end{equation}
and by the definition
of $\hat{\xi}$, the fields are uncorrelated at distance $r$. Therefore
$\langle n(\vec{x})\vec{\nabla}\theta(\vec{x}')
\rangle=0$, and consequently, $N_C(r)=0$.
Thus $N_C(r)$ must be a decreasing function of $r$, as shown in
Fig.~\ref{fig:ncorr} by dashed lines.
Using
Eq.~(\ref{equ:dNCr}) this implies that there is a negative correlation
between the vortices: Wherever there is a vortex, there is likely to
be an antivortex nearby.

The distribution produced by the flux trapping mechanism is very different. As
Eq.~(\ref{equ:cluster}) shows, the vortices tend to form clusters of
radius $\hat{\xi}$, each containing
$N_{\hat{\xi}}$ vortices. Therefore, at short distances, $r\ll\hat{\xi}$,
each new vortex encountered when $r$ is increased typically has the
same sign, and consequently $dN_C/dr>0$. At long distances,
$r\gg\hat{\xi}$, the vortex distribution is the same as the initial
magnetic field distribution,
\begin{equation}
\langle n(0)n(r)\rangle \propto \langle B(0)B(r)\rangle_{\rm equilibrium}.
\end{equation}
In the vacuum approximation (\ref{equ:symmcorr}), the magnetic
field is uncorrelated between different points. Therefore $\langle
n(0)n(r)\rangle=0$, and Eq.~(\ref{equ:dNCr}) shows that
$dN_C/dr=0$. This means that at distances less than $\hat{\xi}$,
$N_C(r)$ is increasing, and near $\hat{\xi}$ it reaches a constant
value. This is shown in Fig.~\ref{fig:ncorr} by solid lines.
Using Eq.~(\ref{equ:dNCr}), we can also say that the vortices
have positive correlations at short distances: Wherever
there is a vortex, there is likely to be another one with the same
sign nearby.

Thus, we have seen that there is a clear, qualitative difference in
the vortex distributions formed by the two mechanisms. In practice,
both mechanisms are believed to 
operate, but by measuring the form of the distribution, we can find
out which one of them is dominant. This has already been done in
numerical simulations, which are discussed in more detail in
Section~\ref{sect:methods},
but it can also be done fairly straightforwardly in superconductor
experiments. This possibility is discussed in Section~\ref{sect:supercond}.

%\newpage

\section{Theoretical methods for studying phase transition dynamics}
\selabel{sect:methods}
\noindent
Ideally, the Kibble-Zurek and flux trapping 
mechanisms could be easily tested numerically by
solving the dynamics of the phase transition in some simple, yet
non-trivial enough quantum field theory. However, exact analytical
solutions for quantum dynamics are possible only in very simple
special cases. Therefore, some approximations are unavoidable, and
many different approaches have been attempted. 

The methods are typically variants of either the Gaussian or the
classical approximation. Examples of the first group are the linear
approximation, in which interactions are neglected altogether and which
corresponds to tree-level perturbation theory, and the Hartree
approximation, which is essentially a tadpole resummation of the
perturbative expansion. The advantage of these approaches is that we
can, at least in a certain sense, 
keep the full quantum mechanical nature of
the system. The price we have to pay is that the self-interaction of
the field is not described correctly, and therefore the approximations
can only be valid at early times. For the same reason, the Gaussian
approximations do not describe the thermal equilibrium state correctly near
the transition
and can therefore at best reproduce Eq.~(\ref{equ:zurekpred0}) with
the mean-field exponents rather than with the true ones. 
Another weakness of the approach is that it is
difficult to apply to more complicated theories, such as gauge field
theories.

Classical approximation relies on the observation that in thermal
equilibrium, long-wavelength modes have high occupation numbers and
can therefore be assumed to behave classically. The most common
realization of this idea is the time-dependent Ginzburg-Landau (TDGL)
approximation, which consists of adding  phenomenological noise and damping
terms to the equations of motion. The main advantage of the classical
approximation is that it describes the non-perturbative interactions
of the long-wavelength modes correctly and, therefore, it also gives a
correct description of the equilibrium state. However, it does not
include any quantum effects and it is, after all, a purely
phenomenological theory, which does not necessarily describe correctly
the dynamics of the intended fundamental theory. We should also be
careful with the renormalization of the model, because the classical
field theory has different ultraviolet behaviour from the original
quantum theory.

The hard-thermal-loop (HTL) approximation incorporates one-loop
quantum corrections to the classical theory in order to avoid the
weaknesses of the TDGL approach. 
It requires more resources than the classical approximation but is
believed to approximate both the non-perturbative and the quantum
mechanical aspects of the system correctly at reasonably long
distances and time scales.

\subsection{Linear approximation}
\noindent
The simplest approximation for the non-equilibrium dynamics is to
approximate the system by a free field theory,\cite{Gill:1995ye} 
\ie, to neglect all
interactions. One advantage of this approach is that one can study the
time evolution of the full density matrix and therefore address
issues such as decoherence and classicality.\cite{Lombardo:2001vs}

Let us consider a global O($N$) scalar field theory  with 
the Lagrangian
\begin{equation}
\label{equ:ONlagr}
{\cal L}=\frac{1}{2}\partial_\mu \phi_i\partial^\mu\phi_i
-\frac{m^2}{2}\phi_i\phi_i-\frac{\lambda}{4}\phi_i\phi_i\phi_j\phi_j,
\end{equation}
where $i,j\in\left\{1,\ldots,N\right\}$. In the linear approximation, the
equation of motion for each Fourier mode $\phi_i(k)$ becomes
\begin{equation}
\partial_0^2\phi_i(k)=-(k^2+m^2)\phi_i(k).
\end{equation}
If $m^2<0$, the modes with $k<\sqrt{-m^2}$ are unstable and grow
exponentially. This approximation works in an instantaneous quench
with Gaussian initial conditions until
\begin{equation}
\langle\phi_i\phi_i\rangle=\int \frac{dk}{2\pi}
 |\phi_i(k)|^2\approx  -m^2/\lambda.
\end{equation}

Because the interaction term is neglected altogether, the linear
approximation does not describe topological defects at all, but it still
shows how the long-wavelength modes grow during the early stages of
the time evolution. Therefore, if we assume that the defect
distribution is related to the field spectrum in some particular way,
it can give some information about the topological defects
formed in the transition.
A popular assumption is to identify the zeroes of the field with
topological defects. In the special case of a Gaussian field, every
observable is given by the two-point function $G(x)$, and therefore
the density of zeroes can also be written as\cite{Halperin,Mazenko-Liu} 
\begin{equation}
\label{equ:HML}
\langle n_0\rangle=C\sqrt{\left|\frac{G''(x=0)}{G(x=0)}\right|},
\end{equation}
where the constant $C$ depends on the dimensionality of the space and
on the symmetry group.

In principle, the linear approximation can also be used for a
time-dependent $m^2$, but in practice the quench must be 
instantaneous,\cite{Gill:1995ye} because otherwise
$\langle\phi_i\phi_i\rangle$ reaches $-m^2(t)/\lambda$ immediately and
the approximation breaks down.
Nevertheless, the approach has been generalized to a 
slowly varying $m^2$ by Karra and 
Rivers.\cite{Karra:1996xf,Karra:1997it} In the case of a
dissipationless (underdamped) field,\cite{Karra:1997it} they found
$\hat{\xi}\propto\tau_Q^{1/3}$, and in the 
overdamped case\cite{Karra:1998gn} 
$\hat{\xi}\propto\tau_Q^{1/4}$. Both of these results agree 
with Zurek's mean-field
prediction (\ref{equ:zurekpred}).

Non-instantaneous quenches have also been studied in a stochastic
(1+1)-dimen\-sional theory in the linear 
approximation by Lythe\cite{Lythe1996} and Moro and
Lythe.\cite{Moro1999}
In the overdamped case,\cite{Lythe1996} the density of kinks was
$n\propto\tau_Q^{-1/4}$, which agrees with Zurek's mean-field
prediction (\ref{equ:zurekpred}). In the underdamped
case,\cite{Moro1999} they found a logarithmic correction to Zurek's
result, 
$n\propto\tau_Q^{-1/3}\ln\tau_Q$, and they were also able to calculate
analytically the prefactors of the power law.
On the other hand, Dziarmaga\cite{Dziarmaga1998} has argued
that the $\tau_Q$ dependence of the underdamped case should actually 
be $n\propto\tau_Q^{-1/2}$ in very slow quenches.

\subsection{2PI formalism}
\noindent
One widely used way of going beyond the linear approximation is to
use the two-particle irreducible (2PI) effective 
action.\cite{Cornwall:1974vz,Calzetta:1988cq}
It is a straightforward generalization of the ordinary effective action
to the case of a field correlator. Let us, for instance,
consider the two-point function
$G(x,y)=\langle\phi^\dagger(x)\phi(y)\rangle$, assuming for simplicity
that the expectation value $\langle\phi\rangle$ vanishes. 
We introduce an external
source $K(x,y)$,
\begin{equation}
Z[K]={\rm Tr}\left[\rho\exp\left(\frac{i}{\hbar}
\int dx dy K(x,y)\phi^\dagger(x)\phi(y)\right)\right].
\end{equation}
Then the two-point function is simply given by
\begin{equation}
\label{equ:Gdef}
G(x,y)=-i\hbar\frac{\partial\ln Z[K]}{\partial K(x,y)}.
\end{equation}
The 2PI effective action $\Gamma[G]$ is defined as
a Legendre transform
\begin{equation}
\Gamma[G]=-i\hbar\ln Z[K]-\int dx dy 
G(x,y)K(x,y),
\end{equation}
where $K=K[G]$ satisfies Eq.~(\ref{equ:Gdef}).
It then follows that
\begin{equation}
\frac{\partial\Gamma[G]}{\partial G(x,y)}=-K(x,y),
\end{equation}
and thereby the minimum of $\Gamma$, given by 
$\partial\Gamma[G]/\partial G(x,y)=0$, determines the
value of $G(x,y)$ in the absence of an external source.
This expression is useful because $\Gamma[G]$ can be written
as a sum over all 2PI vacuum diagrams\cite{Cornwall:1974vz}
and can therefore be calculated fairly easily in perturbation theory.

Although the 2PI formalism is in principle exact, in practice
$\Gamma[G]$ can only be calculated approximatively. In
perturbation theory, the 
lowest non-trivial contribution comes from two-loop diagrams.
This two-loop approximation is equivalent to the Gaussian (Hartree) 
approximation.\cite{Cornwall:1974vz,Chang:1975dt}
If we consider again the O($N$) scalar field theory 
(\ref{equ:ONlagr}),
the equation of motion becomes
\begin{equation}
\label{equ:2PIeom}
\partial_0^2G(k)=-\left[k^2+m^2+\left(2+N\right)\lambda
\int \frac{dk'}{2\pi}G(k')
\right]G(k),
\end{equation}
where we have defined $G(k)$ as the Fourier transform of $G(x)$ given by 
$\langle\phi_i(x)\phi_j(y)\rangle=\delta_{ij}G(x-y)$.
As in the linear approximation, we can then use Eq.~(\ref{equ:HML})
to estimate the number density of topological defects.

Because gauge fields lead to technical difficulties within this
approach, it has only been applied to phase
transitions in theories with global
symmetries.\cite{Boyanovsky:1993pf} 
The density of topological
defects formed in an instantaneous quench was calculated by Antunes
and Bettencourt,\cite{Antunes:1997na} who also compared their results
with the linear theory. 
Later, the Hartree approximation was also applied to non-instantaneous
quenches in a two-dimensional theory with a global O(2) symmetry 
by Bowick and Momen,\cite{Bowick:1998kd} 
but they did not calculate the 
density of defects explicitly.
In the case of an expanding universe, a proper numerical test for
Zurek's predictions was carried out by Stephens 
{\it et al}.\cite{Stephens:1999sm} who found
the scaling laws $\hat{\xi}\propto\tau_Q^{0.35}$ for
underdamped and $\hat{\xi}\propto\tau_Q^{0.28}$ for the overdamped cases,
in a reasonable
agreement with Zurek's predictions (\ref{equ:zurekpred}).

Although the Hartree approximation is sometimes called non-perturbative in the
literature, it is only non-perturbative in the same sense as the
calculation of the effective potential (\ref{equ:Veffdef})
 using one-particle irreducible diagrams. 
This means that the critical exponents $\mu$ and $\nu$ 
in Eq.~(\ref{equ:zurekpred0})
produced by this approximation are not the true critical exponents of
the system, but the perturbative ones. Therefore,
the Hartree approximation, or indeed
the 2PI formalism at any finite loop order, {\em cannot} 
give a correct description of the true
dynamics of the phase transition. Nevertheless, it is still a
non-trivial test for Zurek's predictions to
check whether the results of the Hartree approximation agree with the
predicted mean-field behaviour (\ref{equ:zurekpred}).

Being perturbative, the Hartree approximation is also insensitive 
to truly non-perturbative objects
such as topological defects and therefore it
cannot describe the formation of topological defects properly. This is
obvious from Eq.~(\ref{equ:2PIeom}), because the choice of 
the symmetry group O($N$) only affects the strength of the
effective coupling between
different modes, whereas in reality the dynamics at different $N$ are
very different: $N=1$ gives rise to domain
walls, $N=2$ to vortices and $N=3$ to monopoles.
After all, the Hartree approximation is more or less 
equivalent to the leading term
in the large-$N$ approximation, and there are no topological
defects in the large-$N$ limit. 
Another flaw in this approximation is
that it does not describe scattering of different Fourier modes
properly, because each mode only interacts with the average of all the
other modes. 

The reason why Eq.~(\ref{equ:2PIeom}) has such a simple form is that
all the fundamental fields in the theory are related by a symmetry and
therefore they all have the same two-point function. In more realistic
theories, the equation of motion becomes more complicated, and
the application of this formalism to gauge field theories is largely
an unsolved problem. 
A further problem is that
especially in gauge theories, even the equilibrium properties become
non-perturbative near the transition and are therefore not correctly
described by the perturbative 2PI effective action.

At two-loop level, there is no difference between classical and
quantum mechanical time evolution. Because of this and the other
problems mentioned above, it is essential to be able to go
beyond the Gaussian approximation. 
At three-loop level,\cite{Berges:2001ur,Aarts:2001qa} scattering of
modes is taken into account, although it must still be assumed to be
weak. In this approximation, some 
differences appear between classical and quantum
dynamics, but the equations of motion still look very similar. 
One possible alternative to the perturbative calculation of the
effective action $\Gamma(G)$ is the large-$N$ 
expansion,\cite{Aarts:2001yn}
but as mentioned above, it cannot describe topological defects.
Other ways of improving the Gaussian approximation have also been 
proposed.\cite{Cheetham:1996nd,Kim:2000xd,Salle:2001hd}

\subsection{Classical approximation}
\selabel{sect:classical}
\noindent
\subsubsection{Basics}
\noindent
If we are willing to sacrifice the quantum mechanical nature of the
system, it is relatively straightforward to study the dynamics in a
fully non-perturbative way by simply solving numerically the classical
equations of motion. In fact, this is believed to be a good
approximation as long as the temperature is high enough, because 
the defect density and other physically
relevant properties of the system after the transition are determined
by the long-wavelength field modes, and they have high occupation numbers in
thermal equilibrium.\cite{Grigoriev:1988bd}

A significant advantage of the classical approach over the 2PI
formalism is that in the case of static equilibrium observables it
is equivalent to dimensional reduction,\cite{Ginsparg:1980ef,Kajantie:1996dw}
\ie, to an approach in which the full four-dimensional path integral
is replaced by an effective three-dimensional one [see
Eq.~(\ref{equ:DR})]. 
Because the
construction is free from infrared problems, it is believed that the
effective theory describes the phase transition correctly. In
particular, it belongs to the same universality class as the full
quantum theory, and therefore the critical exponent $\nu$ in 
Eq.~(\ref{equ:zurekpred0}) has its true value. However, as
discussed in Section~\ref{sect:finiteT}, the universality arguments do not apply
to the time dependence, and therefore the other critical exponent
$\mu$ is not necessarily reproduced correctly.
Nonetheless, the classical simulations provide a highly
non-trivial, non-perturbative test for the mechanisms of defect formation.

The general strategy in classical simulations is to take an
initial field configuration that is in thermal equilibrium and 
follow its time evolution by solving numerically the classical
equations of motion. 
The thermal initial condition is typically prepared with a Monte Carlo
algorithm of some type.\cite{Montvay,Rothe} Perhaps the most popular
choice is to use 
Langevin dynamics,\cite{Parisi:1981ys,Ukawa:1985hr}
which means adding noise and damping terms to the equations of
motion. In the case of an O($N$) symmetric scalar field theory, this
means
\begin{equation}
\partial_0^2\phi_i+\eta\partial_0\phi_i-\vec{\nabla}^2\phi_i-
m^2\phi_i-\lambda\phi_j\phi_j\phi_i=\zeta_i,
\end{equation}
where the white noise term $\zeta$ satisfies 
\begin{equation}
\left\langle\zeta_i(t,\vec{x})\zeta_j(t',\vec{x}')\right\rangle
=2T\eta\delta_{ij}
\delta(t-t')\delta^{(3)}\!\left(\vec{x}-\vec{x}'\right).
\end{equation}
In the limit of infinite time, $t\rightarrow\infty$, and zero time
step,
the probability distribution of the
field configurations approaches the canonical ensemble
\begin{equation}
p[\phi]\propto\exp(-H[\phi]/T),
\end{equation}
where $H[\phi]$ is the Hamiltonian.

Another popular approach is the Metropolis 
algorithm,\cite{Metropolis:1953am} which consists of
making small random changes in the field configuration
$\{\phi_i\}\rightarrow\{\phi_i'\}$ and calculating the corresponding 
changes $\Delta E$ in the total energy. If $\Delta E<0$, the change is
accepted, and if $\Delta E>0$, it is accepted with probability
$\exp(-\Delta E/T)$. When the number of these updates is large,
the probability distribution of the field configurations again
approaches the canonical ensemble. The Metropolis approach is
generally more flexible than the Langevin approach, because the nature
of the small random changes can be chosen to optimize the
thermalization rate. In certain special cases, modifications of the
Metropolis algorithm such as the heat bath 
algorithm,\cite{Creutz:1980zw} in which one degree of freedom 
is brought in contact with a heat bath at a time, and hybrid 
Monte Carlo,\cite{Duane:1987de} in which one uses the Hamiltonian
time evolution to move the system around in the phase space,
are more efficient.

In three dimensions, the energy density of a classical field
theory in equilibrium is ultraviolet divergent, but this does not
constitute a severe problem, because the space and time must be
discretized in any case in order for the numerical solution to be
possible, and this lattice cutoff $\Lambda=1/a$ provides the necessary ultraviolet
regularization. Nevertheless, these extra contributions should be
renormalized, \ie, the bare couplings used in the simulations must be
chosen in such a way that they correspond to the desired infrared
physics. In the limit of very fine lattices, $a\rightarrow 0$, 
we can calculate
perturbatively the counterterms that are sufficient for renormalizing
all static equilibrium observables,\cite{Laine:1995ag,Laine:1998dy}
but in gauge field theories, 
they leave some lattice spacing dependence in the time
evolution.\cite{Bodeker:1995pp,Moore:2000fs} 
In particular, this discrepancy becomes worse in the continuum limit,
and therefore we should not use too fine lattices.

It is quite common to keep phenomenological noise and damping terms in the
equations of motion even during the time evolution of the system, in
order to approximate the coupling to the other degrees of
freedom. This time-dependent Ginzburg-Landau (TDGL) approach 
leads to a
faster thermalization than the Hamiltonian equations of motion, but 
it does not change the nature of the dynamics qualitatively, 
because the interactions lead to thermalization in any case.

One advantage of classical simulations is that we can locate the
defects directly in the field configurations and do not have to rely
on the Halperin-Mazenko-Liu approximation (\ref{equ:HML}).
In a global U(1) theory, the lattice winding number is a
straightforward analogue of the continuum one (\ref{equ:cont_wind}). 
For each link $(\vec{x},\vec{x}+\hat{i})$, we define
\begin{equation}
\label{equ:global_Ni}
\Delta\theta_{(\vec{x},\vec{x}+\hat{i})}=
\left[
\theta(\vec{x}+\hat{i})-\theta(\vec{x})\right]_{\pi},
\end{equation}
where the notation $[\ldots]_\pi$ indicates that
 the difference is calculated in such a way that it always lies
between $-\pi$ and $\pi$. This is equivalent to the geodesic rule.
The winding number of a plaquette is then simply
 [cf.~Eq.~(\ref{equ:cont_wind})]
\begin{equation}
\label{equ:global_Nij}
N_{ij}(\vec{x})=\frac{1}{2\pi}\left(
\Delta\theta_{(\vec{x},\vec{x}+\hat{i})}
+\Delta\theta_{(\vec{x}+\hat{i},\vec{x}+\hat{i}+\hat{j})}
-\Delta\theta_{(\vec{x}+\hat{j},\vec{x}+\hat{i}+\hat{j})}
-\Delta\theta_{(\vec{x},\vec{x}+\hat{j})}
\right).
\end{equation}

\subsubsection{Global simulations}
\noindent
Antunes and Bettencourt\cite{Antunes:1997na} 
compared the dynamics of the classical (1+1)-dimen\-sion\-al
scalar field theory in an instantaneous quench
with linear and Hartree approximations. Their
results show that both approximations break down when the back-reaction
becomes important, \ie, when the order parameter 
reaches the minimum of the potential.
Laguna and Zurek\cite{Laguna:1997pv,Laguna:1997sm} studied the
formation of kinks in a linear quench in the same model in the
overdamped case. They found a power law dependence
$n\propto\hat{\xi}^{-1}\propto\tau_Q^{-0.28\pm0.02}$, 
which agrees with the mean-field Zurek prediction (\ref{equ:zurekpred}).
Later,\cite{Laguna:1998cf} they investigated different cases that
interpolate between the overdamped and underdamped dynamics and found
power-law exponents ranging from $-0.23\pm0.01$ (overdamped) to
$-0.33\pm0.01$ (underdamped). 
Moro and Lythe\cite{Moro1999} also simulated the same system 
and confirmed their 
analytical prediction for the prefactor of the power law.

The reason why the above results are consistent with the mean-field
expectations is that there is actually no phase transition at all in
any (1+1)-dimensional classical model, and therefore no critical
behaviour either. This also means that we cannot unambiguously
measure the final defect density. In order to test the scenarios of
defect formation properly, we have to go to higher dimensions.

Antunes {\it et al.}\cite{Antunes:1999rz} studied the formation
of vortices in a (3+1)-dimensional scalar field theory with a global
O(2) symmetry. They used the TDGL approach and, instead of varying
the mass parameter $m^2$, they changed the amplitude of the noise
term, which effectively means changing the temperature of the heat
bath to which the system is coupled. The final density of vortices
per unit area had a power-law dependence on the quench rate
$n\propto\hat{\xi}^{-2}\propto
\tau_Q^{-0.4982\pm0.079}$. This is compatible with the overdamped
mean-field prediction (\ref{equ:zurekpred}), although
the classical simulation should experience the true critical behaviour
with non-mean-field critical exponents.
This is an indication that the freeze-out of the order parameter field
took place before the system entered the critical region.

In their simulations within the same model,
Bettencourt {\it et al.}\cite{Bettencourt:2000jv}
first carried out a linear temperature quench but then
heated the system slightly
above the critical temperature for a short time.
They found that the original vortex
network survives this period of symmetry restoration, which supports Zurek's
idea that the defect distribution is determined by degrees of freedom 
whose dynamics are very slow near the transition
even in the symmetric phase.

The model was also studied by Bowick {\it et al.}\cite{Bowick:2001xg}
who used a time-dependent mass term to start the phase transition.
They concentrated on the evolution of the correlation lengths and
showed that a naive application of Eq.~(21) overestimates the number
of defects by a factor of $2\ldots 4$, because the correlation lengths
keep on growing after the transition.

Stephens\cite{Stephens:2000qv} used the TDGL approach
to study the phase transition in the (2+1)-dimensional
scalar field theory with a global O(3) symmetry.
The topological defects in this model are textures, whose 
energy is independent of their size in the classical
vacuum. Therefore not only the density but also the typical size of
these objects is determined by the phase transition dynamics.
The results show that the typical distance of the textures behaves as
$L_{\rm sep}\propto\tau_Q^{0.39\pm0.02}$ and the size as
$L_{\rm w}\propto\tau_Q^{0.46\pm0.04}$, whereas Zurek's prediction 
(\ref{equ:zurekpred}) for the
freeze-out scale would have been $\hat{\xi}\propto\tau_Q^{0.25}$.
However, these length scales evolve dynamically after the transition, 
which may explain the discrepancy.

\subsubsection{Gauge field theories}
\noindent
One clear advantage of the classical approximation is that it can be
easily applied to models with gauge symmetries. 
We shall discuss only the Abelian Higgs model here, but in principle
similar techniques can and have been applied to non-Abelian 
theories.\cite{Ambjorn:1995xm}
The classical equation of motion for the gauge field is
\begin{equation}
D_\mu F^{\mu\nu}=j^\nu,
\end{equation}
where $j^\nu=-2e{\rm Im}\phi^*D^\nu\phi$ is the electric current.
It is most convenient
to use the temporal gauge $A_0=0$, because that simplifies the
equation of motion into
\begin{equation}
\partial_0 \vec{E}=\vec{\nabla}\times\vec{\nabla}\times\vec{A}+\vec{j},
\end{equation}
where $\vec{E}=-\partial_0\vec{A}$ is the electric field.
However, in the temporal gauge we also obtain an
extra constraint, which the field configurations must satisfy and which
is nothing but the Gauss law
\begin{equation}
\vec{\nabla}\cdot\vec{E}=\rho,
\end{equation}
where $\rho$ is the electric charge.

The necessary discretization of this model is 
straightforward.\cite{Moriarty:1988fx} 
There are two alternative formulations, compact
and non-compact, which correspond to gauge groups U(1) and $\mathbb
R$, respectively. 
Because vortices are not absolutely stable in the compact
formulation, we only consider the non-compact
formulation here. The gauge field is described by a real-valued field
$\alpha_\mu=aeA_\mu$, which is defined on links between lattice
sites. 

It is obvious that Eqs.~(\ref{equ:global_Ni}) and
(\ref{equ:global_Nij}) cannot be used in gauge
field theories, since they are not gauge invariant. If the gauge is
not fixed, the phase angles are totally random and uncorrelated, and
therefore it is always trivially true that
$N_{ij}=1/3$.\cite{Kajantie:1998bg}
The solution is to ``relax the geodesic rule'' by replacing the phase
difference in Eq.~(\ref{equ:global_Ni}) with its gauge-covariant
counterpart\cite{Ranft:1983hf,Kajantie:1998bg}
[cf.~Eq.~(\ref{equ:local_delta})]
\begin{equation}
\label{equ:local_Ni}
\Delta_{\rm cov}\theta_{(\vec{x},\vec{x}+\hat{i})}=
\left[
\alpha_i(\vec{x})+
\theta(\vec{x}+\hat{i})-\theta(\vec{x})\right]_{\pi}
-\alpha_i(\vec{x}).
\end{equation}
When this is used in Eq.~(\ref{equ:global_Nij}), the winding number becomes 
gauge invariant, and it is easy to see that it is always an integer.
In fact, it is the natural lattice analogue of the continuum winding number 
$N_W$ defined in Eq.~(\ref{equ:NWdefgauge}).

\subsubsection{Gauge simulations}
\noindent
Yates and Zurek\cite{Yates:1998kx} investigated the phase transition
in the (2+1)-dimensional Abe\-li\-an Higgs model using the TDGL approach.
They measured the number density of vortices after the transition, and
found power laws $n\propto\hat{\xi}^{-2}\propto\tau_Q^{-0.44\pm0.10}$ 
for the
overdamped case and
$n\propto\hat{\xi}^{-2}\propto\tau_Q^{-0.60\pm0.07}$ 
for
the underdamped case. These results agree with the mean-field
prediction (\ref{equ:zurekpred}) based on the Kibble-Zurek mechanism.
The values they used for the gauge coupling constant and temperature,
$e=0.5$ and $T=0.01$, were so low that the flux trapping
mechanism (\ref{equ:HRdens}) does not have significant effects.
Indeed, the typical flux through their whole $512^2$ lattice
is merely $\Phi\approx 4\Phi_0$ according to Eq.~(\ref{equ:fluctPhi}).
They also used Eq.~(\ref{equ:global_Ni}) to define the winding
number, but again because of the low temperature, this did
not affect the results.

The (2+1)-dimensional Abelian Higgs model
was also studied by Ibaceta and Calzetta,\cite{Ibaceta:1999yy} who
used the overdamped TDGL approach. Because they used an instantaneous
quench, they were not able to test the predictions for the defect
density. On the other hand, their aim was to test the applicability of
the linear approximation, and they found that although it eventually
breaks down, it agrees very well at
early times. This is presumably partly because they used Gaussian
initial conditions for the Higgs field rather than thermal ones. 
In their initial conditions, the gauge field
vanished everywhere,
so they could not have observed the flux trapping
mechanism
in any case. Furthermore, they used Eq.~(\ref{equ:global_Ni}) as the
definition for
the winding number.

The only (3+1)-dimensional simulations in a gauge field theory so
far have been carried out by Hindmarsh and 
Rajantie.\cite{Hindmarsh:2000kd,Hindmarsh:2001vp}
In Ref.~\acite{Hindmarsh:2000kd}, they used the classical approximation
without damping and noise terms. In their lattice, one of the dimensions 
was shorter than the other two, in order to stabilize vortices
that wind around that dimension so that they can be counted. They
prepared the initial configuration using a hybrid Monte Carlo
algorithm,\cite{Rothe} and used a time-varying mass term in their
simulations. They used the gauge-invariant definition
(\ref{equ:local_Ni})
for the winding
number to locate the defects in the final state, and found the
power-law $n\propto\tau_Q^{-0.250\pm0.013}$ when the short dimension was 5
in lattice units, and $n\propto\tau_Q^{-0.274\pm0.039}$ when it was
20. These results disagree strongly with Zurek's prediction
(\ref{equ:zurekpred}), which is $n\propto\tau_Q^{-2/3}$. 
On the other hand, they can be explained in terms of
the flux trapping mechanism.

From the viewpoint of vortex dynamics, the system they studied was
effectively two-dimensional and therefore the relevant theoretical
predictions correspond to $D=2$ in Eq.~(\ref{equ:HRdens}), 
\ie,\cite{Hindmarsh:2000kd}
\begin{equation}
\label{equ:HRLz}
n\approx\frac{e}{2\pi}\frac{T^{1/2}}{L_z^{1/2}\hat{\xi}},
\end{equation}
where $L_z$ is the extent of the system in the short dimension. This
predicted dependence on $L_z$ was confirmed in the
simulations.\cite{Hindmarsh:2000kd} Note,
however, that using Eqs.~(\ref{equ:HRhatxi}) and (\ref{equ:HRLz}), we
expect $n\propto\tau_Q^{-0.2}$. This discrepancy is explained by
the failure of the classical approximation to describe the time
evolution of the quantum theory correctly. In particular, Landau
damping (\ref{equ:pertlandau}) 
is very sensitive to the quantum mechanical ultraviolet modes,
and in the classical theory, it actually behaves more 
like\cite{Hindmarsh:2001vp}
$\gamma_L(k)\propto k^{2.1}$. Substituting this into
Eqs.~(\ref{equ:adiab}) and (\ref{equ:HRLz}), we find 
$n\propto\tau_Q^{-0.24}$, in a fair agreement with the measurements.

\begin{figure}
\center
\epsfig{file=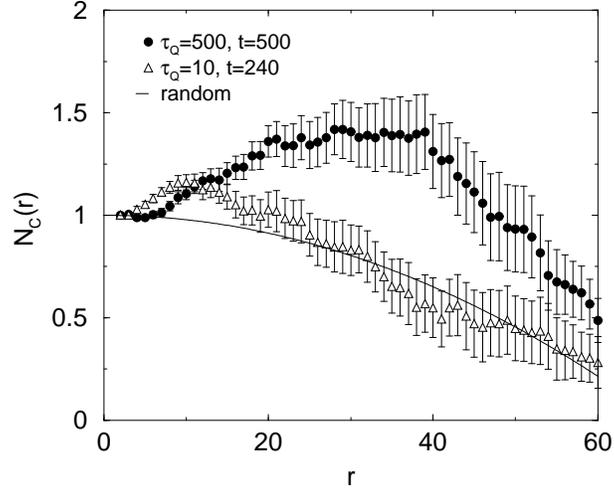,width=8cm}
\flushleft
\caption{
\label{fig:meas_ncorr}
The correlation function $N_C(r)$ [see Eq.~(\ref{equ:NCr})] measured
by Hindmarsh and Rajantie\protect\cite{Hindmarsh:2000kd} in two quenches with
different $\tau_Q$. The solid line illustrates the effect of the
finite system size by showing $N_C(r)$ for a random distribution of
vortices and antivortices. In both runs, $N_C(r)$ is above this curve,
which supports the flux trapping scenario, as discussed in
Section~\ref{sect:spatdist} (see also Fig.~\ref{fig:ncorr}).
(From Ref.~\protect\acite{Hindmarsh:2000kd}.)}
\end{figure}

As discussed in Section~\ref{sect:spatdist}, 
the flux trapping mechanism also predicts a characteristic 
spatial distribution for
the vortices. Hindmarsh and Rajantie\cite{Hindmarsh:2000kd} 
studied that in their
simulations by measuring the function $N_C(r)$ [see Eq.~(\ref{equ:NCr})].
Their results for two different quench rates are shown in
Fig.~\ref{fig:meas_ncorr},
and they display a clear positive correlation at short distances, as
predicted by the flux trapping mechanism.

\begin{figure}
\center
\epsfig{file=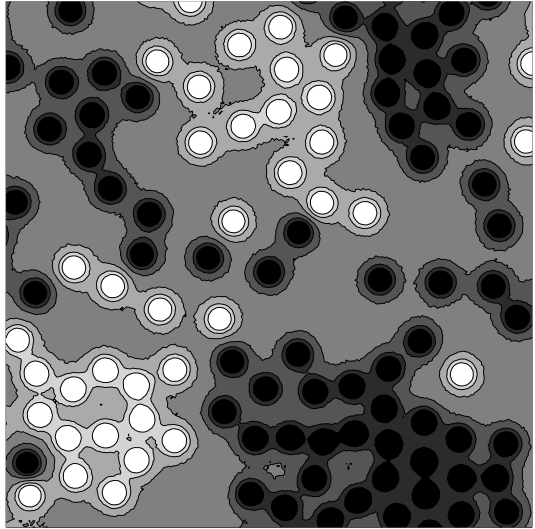,width=7cm}
\flushleft
\caption{
\label{fig:stephens}
An example of the spatial vortex distribution after a phase
transition in the simulations of Stephens 
{\it et al.}\protect\cite{StephensNew}
White and black circles correspond to vortices and antivortices,
respectively. The plot shows clearly the clustering of vortices
discussed in Section~\ref{sect:spatdist}. 
(From Ref.~\protect\acite{StephensNew}, kindly supplied by G.~Stephens.)
}

\end{figure}

More recently, Stephens {\it et al.}\cite{StephensNew}
studied instantaneous phase transitions in the (2+1)-dimensional
Abelian Higgs model using overdamped TDGL simulations. 
They found very strong clustering of vortices, as shown in
Fig.~\ref{fig:stephens}, in agreement with the discussion in
Section~\ref{sect:spatdist}. They also measured the dependence of the
vortex density on the gauge coupling and temperature, and found the
relation $n\approx CeT^{1/2}$, where $C$ is a constant. This confirms
the prediction (\ref{equ:HRdens}) of the flux trapping scenario.

People have also studied inhomogeneous setups with the motivation that
they might be more relevant for actual experimental phase 
transitions.\cite{KibbleInhomog}
Dziarmaga {\it et al.}\cite{Dziarmaga:1998ie} used the TDGL approach
in a global theory in
(1+1) dimensions to study a phase transition which starts from one
end of the system and propagates to the other. They found significant
suppression of defect density.
Aranson {\it et al.}\cite{Aranson1999,Aranson2001} have used
overdamped TDGL
simulations to study a global (3+1)-dimensional U(1) theory in an
inhomogeneous case, in which the system was initial heated up locally
at a single point.

\subsection{Hard-thermal-loop improvement}
\noindent
When discussing the classical approximation in
Section~\ref{sect:classical}, we implicitly assumed that the classical
theory had the same Lagrangian as the full quantum theory it was
supposed to approximate. However, B\"odeker {\it et
al.}\cite{Bodeker:1995pp,Moore:2000fs}
showed that this leads to wrong dynamics, although it reproduces
the static equilibrium properties correctly, once simple renormalization
counterterms are included.

The reason for this failure is that 
the classical approximation is only valid for long-wavelength modes.
 Therefore, the classical
theory must be interpreted as an effective theory in the Wilsonian
sense, and in principle, its Lagrangian should contain all the terms
that are compatible with the symmetries of the system. In vacuum,
these terms are strongly constrained by the Lorentz invariance, but at
a non-zero temperature, Lorentz invariance is broken by the rest frame
of the thermal background. 
Therefore many more terms are allowed in the effective Lagrangian and
restricting it to its original form is not justified.

Fortunately, the difference between the quantum and classical theories
appears only at very high momenta, and these ultraviolet degrees of
freedom are perturbative.
Therefore we can calculate the necessary corrections to the classical
Lagrangian perturbatively. In a global scalar theory, we only obtain
a simple mass counterterm,\cite{Aarts:1997qi} 
but in gauge field theories, the correction
is a complicated non-local object.\cite{Pisarski:1989vd}
In the Abelian Higgs model, this {\em hard-thermal-loop} improved
effective Lagrangian is\cite{Kraemmer:1995az}
\begin{eqnarray}
{\cal L}_{\rm HTL}&=&-\frac{1}{4}F_{\mu\nu}F^{\mu\nu}
-\frac{1}{4}m_D^2
\int\frac{d\Omega}{4\pi}F^{\mu\alpha}
\frac{v_\alpha v^\beta}{(v\cdot\partial)^2} F_{\mu\beta}\nonumber\\
&&+|D_\mu\phi|^2
-m_T^2|\phi|^2-\lambda|\phi|^4,
\label{equ:lagrHTL}
\end{eqnarray}
where $m_T^2=m^2+(e^2/4+\lambda/3)T^2$, and the integration 
is taken over the unit
sphere of velocities $v=(1,\vec{v})$, $\vec{v}^2=1$.
The Debye screening mass has the value $m_D^2=\frac{1}{3}e^2T^2$.
We have here ignored a similar, non-physical, contribution that arises
from ultraviolet lattice modes, and which should in principle be
subtracted.\cite{Nauta:2000cm} 
Therefore we must assume that the lattice spacing is long enough,
$a\gg T^{-1}$, so that this
contribution is negligible.
Because the perturbative calculation that gave Eq.~(\ref{equ:lagrHTL})
involved only hard, ultraviolet modes, it is believed to be reliable
as long as the coupling constants are small.\cite{Rajantie:1999mp} 

In principle, we can simulate the dynamics of the theory by solving
numerically the equations of motion derived from
Eq.~(\ref{equ:lagrHTL}).
However, because the extra term is non-local 
this requires keeping all the previous time
steps in the memory. 
Instead, it is much more convenient to introduce extra degrees
of freedom, which have local interactions, and which reproduce the
same effective term. There are different ways of doing this: The extra
degrees of freedom can either be point particles\cite{Moore:1998sn} or
fields.\cite{Blaizot:1999xk,Rajantie:1999mp,Bodeker:2000gx}

The HTL approach has been used to study equilibrium quantities such as
the rate of baryon number
violation in the symmetric phase of the electroweak
theory.\cite{Moore:1998sn,Bodeker:2000gx}
Similarly, Hindmarsh and
Rajantie\cite{Hindmarsh:2001vp}
 used it in the
Abelian Higgs model to measure the real-time correlator of the
magnetic field at the transition point. They found that 
within the range of wavelengths they studied,
the
time scale that dominates the dynamics of the long-wavelength magnetic
fields is determined by the Landau damping rate $\gamma_L$ 
and that at
intermediate wavelengths it agrees with the
perturbative result (\ref{equ:pertlandau}). 
Because $\gamma_L$ is very sensitive to quantum effects --- they
also measured the rate without the HTL correction and obtained a
very different result --- this is a very strong indication that
the HTL approach really approximates the full quantum dynamics well.
Nonetheless, because the approximation ignores the scattering of hard
modes, it eventually breaks down at very long wavelengths, $k\lsim
e^4T$, and very long times.

The HTL approximation can also be used in non-equilibrium settings
provided
that the ultraviolet modes are still in equilibrium, which is the case
in thermal phase transitions.\cite{Rajantie:1999mp} 
Hindmarsh and Rajantie\cite{Hindmarsh:2001vp}
used this approach to study vortex formation in the phase transition
of the Abelian Higgs model, and found $n\propto\tau_Q^{-0.201\pm0.015}$.
This differs from
the prediction of the Kibble-Zurek mechanism (\ref{equ:zurekpred}) which is
$n\propto\hat{\xi}^{-2}\propto\tau_Q^{-0.5\ldots-0.66}$,
but agrees very well with the flux trapping predictions
(\ref{equ:HRhatxi}) and 
(\ref{equ:HRLz}).

%\newpage
\section{Experiments}
\selabel{sect:expt}
\noindent
The similarity of defect formation in condensed matter systems and
cosmology was already noted by Zeldovich {\it et
al.}\cite{Zeldovich:1974uw} and
Kibble,\cite{Kibble:1976sj,Kibble:1980mv} but the first concrete
proposal of utilizing this correspondence to test the cosmological
scenarios in experiments was made 
by Zurek.\cite{Zurek:1985qw,Zurek:1993ek}
He suggested an experiment in which a pressure quench would be used to cause
a phase transition in $^4$He and which was later carried out by Hendry
{\it et al.}\cite{Hendry}

Later on, other condensed matter systems have been used to study
defect formation as well. Most of them, liquid
crystals,\cite{TurokNature,Bowick:1994rz} $^4$He\cite{Hendry,Dodd} 
and $^3$He-B\cite{Bauerle,Ruutu:1996qz} are systems with global
symmetries, and therefore the applicable theoretical scenario is the
Kibble-Zurek scenario. The same applies to recent studies of
non-linear optical systems\cite{Ducci1999} and convection in 
fluids,\cite{Casado2001}
as well as to the proposed experiments with atomic Bose-Einstein
condensates.\cite{Anglin:1999pm}

The only exception are superconductors.\cite{Tilley} 
The order parameter, the
Cooper pair, is electrically charged, and therefore the symmetry that
is broken is a local symmetry. This means that the flux trapping 
mechanism should
form vortices in addition to the Kibble-Zurek mechanism. So far, only two
superconductor experiments have been carried out,\cite{Carmi,Carmi_Jos}
and they haven't been able to give conclusive results, but the rapid
progress in experimental techniques in the recent years suggests that more
detailed experiments will soon be possible.

\subsection{Liquid crystals}
\selabel{sect:LCD}
\noindent
Liquid crystals\cite{deGennes}
are perhaps the simplest condensed matter systems in which
formation of topological defects can be studied. The phase transition
between isotropic and nematic phases
takes place near the room temperature, and the defects can be seen with
an optical microscope. A disadvantage of liquid crystals is that the
transition is of first order and therefore it cannot really probe the
Kibble-Zurek mechanism, which applies to continuous phase transitions.

A nematic liquid crystal consists of rod-like molecules, whose
orientation acts as the order parameter.
It is an
unoriented three-vector and is commonly
denoted by $\vec{n}$. The vacuum manifold is $S^2/{\mathbb Z}_2$ and
has non-trivial first, second and third homotopy groups. Therefore,
the system has three types of topological defects: vortex lines
(strings), monopoles and textures.

The first experimental study of defect formation was carried out by
Chuang {\it et al.}\cite{TurokNature}
and concentrated mainly on the dynamics of the string network after
the transition from the isotropic to the nematic phase.
Bowick {\it et al.}\cite{Bowick:1994rz} paid more attention to the
formation of the defects, and carried out a quantitative comparison
with the Kibble mechanism. Because the phase transition is of
first order, the number density of vortex lines in the
nematic phase should be given by Eq.~(\ref{equ:kibbledens}). 
They counted the number of bubbles, compared it with the number
of vortices and found a good agreement with the theoretical
prediction.

As discussed in Section~\ref{sect:spatdist}, the spatial distribution of
defects is a very convenient way to test the mechanisms for defect
formation. Because the symmetry broken in the isotropic-nematic phase
transition is global, the relevant theoretical scenario is the Kibble
mechanism, and it predicts strong negative correlations between
defects. These correlations were studied experimentally by Digal {\it
et al.}\cite{Digal:1998ak} They measured the net number of vortices
$N_W(A)$ through a loop of area $A$. On average, this is obviously
zero, because vortices of either sign have the same probability, but
the width $\sigma$ of the distribution depends on the correlations of the
vortices. 
Digal {\it et al.} assumed a power-law behaviour
\begin{equation}
\sigma=CN^\nu,
\end{equation}
where $C$ and $\nu$ are free parameters. 
They used the total number of vortices $N$ inside the loop
instead of $A$, because it separates the effects
of the vortex correlations from the relation between $N$ and $A$.

The Kibble mechanism predicts $\nu=1/4$, but if the vortices 
were randomly distributed, the exponent
would be $\nu=1/2$. The measured value for the exponent was
$\nu=0.26\pm0.11$, in a good agreement with the Kibble mechanism.
Moreover,
Digal {\it et al.} measured the value of the
prefactor $C=0.76\pm0.21$, which
also agrees well with the theoretical prediction $C\approx 0.71$, 
calculated using the simplifying assumption that the bubble nucleation
sites form a square lattice.

\subsection{Superfluids}
\noindent
At low enough temperatures, helium becomes a superfluid,\cite{Tilley} 
and this
transition can also be described as a spontaneous breakdown of a
global symmetry. It has the advantage over liquid crystals that
superfluidity is a quantum phenomenon, and therefore the experiments
can really probe the behaviour of a quantum field theory rather than a
classical one.

In the $^4$He isotope, the phase transition to the superfluid phase
takes place at the critical temperature $T_\lambda\approx 2.18$~K.
In the Ginzburg-Pitaevskii\cite{GinzPit} 
picture of the transition, the order parameter
$\psi$ is the quantum mechanical wave function of the $^4$He atoms.
In practice, $\psi$ is a complex scalar field, and the theory is
invariant under global U(1) transformations.
In the superfluid phase, the $^4$He atoms form a Bose condensate,
which is signalled by a non-zero vacuum expectation value of $\psi$. 
This breaks the U(1) symmetry, and as discussed in
Section~\ref{sect:Topo}, leads to the existence of vortex line
solutions.

Vortex formation in a $^4$He phase transition was first studied
experimentally by Hendry {\it et al.}\cite{Hendry} They had a
small volume of $^4$He in a container at a temperature slightly 
above $T_\lambda$,
and instead of cooling the system, they expanded the container
rapidly so that the pressure decreased. This increases $T_\lambda$,
and therefore the system underwent a phase transition into the
superfluid phase. They counted the vortices formed in the
transition by measuring the attenuation of the second sound in the
liquid after the transition. The results show that vortices were
indeed formed in the transition, but this was later attributed to
hydrodynamic effects that arose from non-idealities in the
experimental setup rather than to
the Kibble-Zurek mechanism.\cite{Gill1996}

A more careful study was carried out later by Dodd {\it et
al.}\cite{Dodd,Dodd2} who eliminated the most significant sources for
hydrodynamic vortex formation. Surprisingly, they did not find any
evidence for vortex formation in the transition. They suggested that
the vortices may have decayed faster than expected.
This idea was made more precise by Karra and
Rivers,\cite{Karra:1998gn}
who argued that the Ginzburg temperature $T_G$ [see
Eq.~(\ref{equ:Tginzburg})], above which thermal fluctuations
are still able to restore the symmetry within a correlated region,
is well below $T_\lambda$ in $^4$He, and in fact, the
final state was still above $T_G$ 
in the experiment.
Therefore, the vortices never became well defined classical
objects and were washed out by thermal fluctuations before being observed.

The other helium isotope $^3$He also displays superfluidity at low
temperatures,\cite{Tilley}
but many details are very different. Because $^3$He
atoms are fermions, they can only condensate if they form bosonic
Cooper pairs. Therefore, superfluidity needs millikelvin temperatures,
much lower than in $^4$He. Furthermore, the Cooper pairs may have
a non-trivial spin ${\bf S}$ and angular momentum ${\bf L}$, 
which means that the order
parameter is more complicated than simply a complex scalar field.
Therefore, there are different superfluid
phases: the A phase, which is only present in a narrow range of
temperatures and under high pressures or if a strong external
magnetic field is applied, and the B phase. In an external magnetic field,
a third superfluid phase, known as the A1 phase, is also possible.

In all the existing experiments, the transitions 
have been between the normal and the B phase, and therefore we shall
only concentrate on that. In the B phase, the total angular momentum
${\bf J}={\bf L}+{\bf S}$ vanishes. The symmetry breaking structure 
is\cite{Tilley}
\begin{equation}
{\rm SO(3)}_{\bf L}\times {\rm SO(3)}_{\bf S}\times {\rm U(1)}_\phi
\rightarrow
{\rm SO(3)}_{{\bf L}+{\bf S}}.
\end{equation}
This allows two different types of vortices, oriented {\it mass vortices} and
unoriented {\it spin vortices}.\cite{VolovikMineev}
The latter ones are bound to soliton sheets and are therefore
unstable.

Vortex formation in $^3$He has been studied in two 
experiments.\cite{Bauerle,Ruutu:1996qz} They both used
external neutron sources to heat up the system locally. When a
neutron hits a $^3$He nucleus, it may be captured in the reaction
\begin{equation}
{\rm n}+^3\!{\rm He} \rightarrow ^3\!{\rm H} + {\rm p}.
\end{equation}
This releases 764 keV of energy, which heats up 
a ``hot spot'' of radius $\approx 30~\mu$m, and inside this hot spot
the symmetry is locally restored. The hot spot cools
rapidly in about $1~\mu$s, and undergoes a phase transition back to the
superfluid phase. According to the Kibble-Zurek scenario, a tangle of vortices
is formed in this process.

In the Helsinki experiment by Ruutu {\it et al.}\cite{Ruutu:1996qz} 
the cryostat that
contained the helium was rotating with a velocity lower than the
critical velocity of the superfluid. This rotation expands those 
vortex loops that are larger than a certain critical size. Eventually, they
straighten out and move to the centre of the container. 
Using NMR spectroscopy, Ruutu {\it et al.} were able to count the
vortices one by one and thereby obtained a very precise measurement for
the number of vortices formed in the transition. 

In the Grenoble experiment carried out by B\"auerle {\it el
al.}\cite{Bauerle} the number of vortices formed in the transition
was inferred using a vibrating 
superconducting wire, which measures the amount of
heat deposited in the liquid very accurately. If vortices are formed,
this amount is less than the total 764~keV released in the neutron
capture event, because some of the energy is stored in the vortices.
The experiment was carried out at a temperature well below $T_c$ so
that the lifetime of the vortices was much longer than the time needed
for the measurement.

In both $^3$He experiments, the measured number of vortices was in
a reasonable agreement with the prediction of the Kibble-Zurek
mechanism.
In experiments of this type,
the quench timescale $\tau_Q$ is not a free parameter, because it is
determined
by the equilibration rate of the system after a neutron capture event,
and therefore it is not possible to study the dependence of the
vortex number on $\tau_Q$. Nevertheless the results show beyond doubt
that vortices are formed in the transition, in contrast to $^4$He, 
which
may be explained by the fact that in $^3$He, $T_G$ is very close to
$T_c$.\cite{Karra:1998gn}

In the above experiments, the observed vortices were mass vortices,
but more recently, Eltsov {\it et al.}\cite{Eltsov:2000ke} have
also reported an observation of composite {\it spin-mass vortices} after a
phase transition to the B phase.

Another type of a superfluid experiment was also suggested by
Zurek.\cite{Zurek:1985qw} He pointed out that if the superfluid
container is an annulus instead of a cylinder, the ``vortices'' that
are formed in the transition pass through the hole. Because
their cores are outside the superfluid, they cost very
little energy and are therefore very stable. 
They can be detected as a non-zero angular velocity of the
superfluid around the annulus. However, this experiment has not been
carried out yet.

\subsection{Other global systems}
\noindent
The liquid crystal and superfluid experiments were not able to measure
the dependence of the vortex number on $\tau_Q$ and therefore they
could not test the Kibble-Zurek mechanism quantitatively.
However, there have recently been reports of other studies in which this
has been done.\cite{Ducci1999,Casado2001}

The first such experiment 
was carried out by Ducci {\it et al.}\cite{Ducci1999}
in a non-linear optical system.
They studied a liquid crystal light valve illuminated by
a laser beam and inserted in a feedback loop. When the intensity of
the incident light exceeds a threshold value, an intensity patterns of
standing rolls forms. The authors changed the intensity of the light 
at a finite
rate characterized by $\tau_Q$, recorded the intensity pattern and
counted the defects.
They found a power-law dependence $n\propto\tau_Q^{-0.50\pm0.04}$, in
a very good agreement with Zurek's prediction (\ref{equ:zurekpred}).

Casado {\it et al.}\cite{Casado2001} studied the breaking of a global
symmetry in the B\'enard-Marangoni
conduction-convection transition. They heated up the bottom of a
cylindrical container with a  layer of silicone oil in it.
Below a certain 
critical temperature $T_c$, the fluid is in
a homogeneous conduction state, but above $T_c$, the symmetry is
broken by a hexagonal array of convection lines where hot fluid flows
upwards.
Casado {\it et al.} captured the image with a CCD camera and 
located the defects in the array structure. They
used different rates of increasing the temperature at the bottom of
the container, characterized by the quench time scale $\tau_Q$.
The number of defects 
had a power-law dependence on $\tau_Q$, with exponent ranging from
$-0.45$ to $-0.25$ depending on the viscosity of the fluid.

\subsection{Superconductors}
\selabel{sect:supercond}
\noindent
All the experiments discussed above are systems with broken global
symmetries, whereas local gauge symmetries are more relevant for
particle physics and cosmology.
Therefore, it has been appreciated for a long 
time\cite{Rudaz:1993wy,Zurek:1996sj,Rudaz:1999ra} that 
defect formation should also be studied experimentally in
type-II superconductors.

In superconductors,\cite{Tilley} 
the order parameter is a scalar field $\psi$, which
describes the Cooper pairs. The field is charged under the U(1) gauge
group
of
electrodynamics, because the Cooper pairs have an electric charge of
$-2e$. In the superconducting phase, the Cooper pairs condense, and this
leads to the Meissner effect.
The equilibrium properties of superconductors near the phase
transition are described by the Ginzburg-Landau theory, which is
very similar to the Abelian Higgs model (\ref{equ:AHdef}).
Although the details of the dynamics are presumably quite different,
the considerations that led to the flux trapping mechanism (see
Section~\ref{sect:HRmech}) are still valid, and therefore the
mechanism can be tested in superconductor experiments.

In principle, the required experimental setup is relatively simple. A
superconducting film is cooled through the phase transition from
the normal phase into the superconducting phase, and the vortices
formed in the transition are detected by measuring their magnetic
field.

In practice, there are many difficulties in this kind of an
experiment. A slightly simpler setup\cite{Zurek:1996sj} 
involves a superconducting loop, which is quenched through the
phase transition. In this kind of a system, the ``vortices''
show up as a non-zero magnetic flux through the loop. The Kibble-Zurek
mechanism
predicts $N_{\rm KZ}\approx(L/\hat{\xi})^{1/2}$ flux quanta,
where $L$ is the circumference of the loop, and the magnetic field
trapped according to the flux trapping mechanism consists of 
$N_{\rm trap}\approx
(e^2TL)^{1/2}$ flux quanta, 
independently of the quench rate, provided that $\hat{\xi}<L$.

An experiment that was essentially like this was carried out by Carmi
{\it et al.}\cite{Carmi_Jos} They used a 1 cm$^2$ loop of
YBa$_2$Cu$_3$O$_{7-\delta}$ superconductor built from 214 Josephson
junctions. A light beam was used to heat the loop above $T_c\approx
90$~K, and when the light was switched off, a thermal link to a heat
bath cooled the loop through the phase transition. The cooling rates
varied from $0.3$ to $20~{\rm K/s}$. They measured the flux through
the loop after the transition using a superconducting quantum
interference device (SQUID) magnetometer placed at a
distance of 1 mm from the loop, and found that its was zero on average
and the distribution had a standard deviation of 
$\sigma_{\rm exp}=7.4\pm 0.7$ flux
quanta, independently of the cooling rate.

In this setup, the Kibble-Zurek mechanism predicts that the phase of the
order parameter is independent in each of the $N=214$ segments, and
therefore the resulting standard deviation should be roughly
$\sigma_{\rm KZ}=C\sqrt{N}\approx 14.6C$, where $C$ is 
some constant of order one. This
result agrees well with the observations, and is indeed independent of
the quench rate, as observed. The contribution predicted by 
the flux trapping 
mechanism
is\cite{Hindmarsh:2000kd} $\sigma_{\rm trap}\approx 4$ flux
quanta, and therefore the experiment cannot really distinguish between
the two mechanisms.

What simplifies the superconductor loop experiment is that the
``vortices'' are formed outside the superconductor itself. This,
however, also means that the experiment does not really probe the
dynamics inside the superconductor, and indeed in the experiment by Carmi
{\it et al.}\cite{Carmi_Jos} the numbers of vortices predicted by the 
Kibble-Zurek and flux trapping 
mechanisms are both independent of the quench rate.

Carmi and Polturak\cite{Carmi} also carried out a similar 
experiment with a superconducting YBa$_2$Cu$_3$O$_{7-\delta}$ film of
size 1 cm$^2$. Again, they heated the film with light, switched the
light off and measured the resulting magnetic flux through the film 
when the system had
reached the superconducting phase. They estimated that the sensitivity
of their measurement was around 20 flux quanta, and they could not
find any evidence of vortices being formed. They compared this
with their own theoretical estimate, which was around 10000, and
concluded that there is a clear discrepancy. However, their
theoretical picture was rather different from the Kibble-Zurek or 
flux trapping scenarios,
which predict much fewer vortices, of order 100 and 1,
respectively.\cite{Zurek:2000ym,Hindmarsh:2000kd}

In the film experiment,\cite{Carmi} the Kibble-Zurek prediction depends on the
quench rate, so one could test the scenario by measuring this
dependence. However, the prediction also depends on details of the
dynamics that have not yet been properly understood, and
therefore it is not a very robust test. 
Moreover, the flux trapping scenario becomes slightly more complicated
because the magnetic field extends outside the two-dimensional
film.\cite{Hindmarsh:2000kd,Rajantie:2001na} 
For this reason, simple two-dimensional simulations
such as those in Ref.~\acite{StephensNew} do not describe
the transition correctly.

On the other hand, as pointed out in Section~\ref{sect:spatdist}, the two
scenarios predict very characteristic correlations between the
produced vortices, and therefore a measurement of the spatial vortex
distribution is the best way of distinguishing between the
mechanisms. Unfortunately, this was not possible in the experiments
carried out by Carmi {\it et
al.}\cite{Carmi,Carmi_Jos}
because they only measured the total magnetic flux.

In order to measure the spatial vortex distribution, one has to face
several technical problems. First of all, one has to be able to
identify individual vortices, which means a sensitivity to fluxes at a
level well below one flux quantum. Secondly, the spatial resolution
must be high enough so that one can see the correlations, and
finally, the measurement must be fast enough so that the vortices do
not have time to annihilate. There are several different methods
available,\cite{Bending} all of which have their strengths and weaknesses. 
Scanning SQUID microscopy has a very high sensitivity and reasonably
high spatial resolution, but the measurements take minutes. 
Real-time imaging at video rate with a sensitivity to a single flux
quantum
has been achieved using
Lorentz microscopy\cite{Harada} and magneto-optical imaging.\cite{Goa}

The dynamics of the vortices after the transition are complicated by
the impurities in the superconductor film, which tend to pin vortices
to them. This effect can also be used to simplify the measurement by
constructing a regular array of pinning sites from submicron holes
(antidots).\cite{Moshchalkov} When the vortices get pinned to the
antidots, their spatial distribution is stabilized and can be measured
with slower techniques such as scanning SQUID microscopy and scanning
Hall probe microscopy.

A further type of an experiment has been proposed by 
Kavoussanaki {\it et al.}\cite{Kavoussanaki:2000tj} who suggested
using an annular Josephson junction consisting 
of two superconducting loops
separated by a thin layer of  insulator. 
This system is described by the sine-Gordon theory, whose field is
the phase angle difference of the order parameter fields in 
   the two loops. A soliton of the
sine-Gordon model corresponds to a case with different magnetic fluxes
through the two loops, and the Kibble-Zurek mechanism predicts
that these solitons are formed in the transition.

%\newpage

\section{Conclusions}
\noindent
In this article, we have discussed the formation of topological
defects in field theory phase transitions. In particular, we
have concentrated on the differences between gauge field theories and
theories with global symmetries. 

While the Kibble-Zurek scenario,\cite{Kibble:1976sj,Zurek:1985qw} 
which applies to theories with global
symmetries, has been around for decades and has been tested both
in numerical simulations and in
various condensed matter experiments, the
same phenomenon in gauge field theories has remained poorly understood
until very recently.
However, in the last couple of years, the role of magnetic field has
been clarified and a 
theoretical picture 
of defect formation based on the trapping of magnetic flux
has emerged.\cite{Hindmarsh:2000kd}
Both the Kibble-Zurek and the flux trapping mechanisms are based on
a freeze-out of long-wavelength modes that are too slow to
adapt to the change of the external parameters. It is noteworthy that
very similar phenomena are expected to take place in heavy-ion
collisions.\cite{Berdnikov:2000ph,Boyanovsky:2001nt}

We have reviewed the basic predictions of the Kibble-Zurek and flux
trapping scenarios. The numbers of vortices formed by these two
mechanisms depend on the temperature, the rate of the phase transition
and other parameters in different ways. Furthermore, the mechanisms lead to
very different spatial distributions of vortices: While the
Kibble-Zurek mechanism predicts negative correlations between the
defects, the correlations are positive in the flux trapping scenario.
This means that flux trapping typically forms clusters of vortices with
equal sign, and depending on the parameters, these clusters can be large.

These predictions have been confirmed in numerical simulations, but
they can also be tested in superconductor experiments, which are made
possible by the recent developments in experimental techniques. This
has been demonstrated in the recent pioneering
experiments,\cite{Carmi,Carmi_Jos}
and there is little doubt that proper
experimental tests will be carried out in the near future.

These experiments will provide us with valuable information about the
non-equilibrium dynamics of gauge field theories during phase
transitions. This is extremely important for particle physics and
cosmology, as well, because once the behaviour in a simpler system
such as a superconductor is understood, similar theoretical techniques
can be applied to heavy ion collisions and cosmological phase
transitions.

%\newpage
\nonumsection{Acknowledgements}
\noindent
The author wishes to thank 
Gert Aarts, Nuno Antunes,
Dan Cormier, Anne-Christine Davis,
Tom Girard, Mark Hindmarsh, Tom Kibble, Ray Rivers, 
Greg Stephens, Neil Turok and Wojciech Zurek
for useful discussions and correspondence. This work was supported by PPARC, 
the ESF programme Cosmology in the Laboratory (COSLAB) and the UK
Thermal Field Theory network.

\nonumsection{References}


\noindent

\begin{thebibliography}{000}

%\cite{Kibble:1976sj}
\bibitem{Kibble:1976sj}
T.~W.~B.~Kibble,
%``Topology Of Cosmic Domains And Strings,''
J.\ Phys.\ A {\bibbf 9} (1976) 1387.
%%CITATION = JPAGB,A9,1387;%%

\bibitem{VilenkinShellard}
A.~Vilenkin and E.P.S.~Shellard,
{\bibit Cosmic Strings and Other Topological Defects}
(Cambridge University Press, Cambridge, 1994).

%\cite{Berdnikov:2000ph}
\bibitem{Berdnikov:2000ph}
B.~Berdnikov and K.~Rajagopal,
%``Slowing out of equilibrium near the QCD critical point,''
Phys.\ Rev.\ D {\bibbf 61} (2000) 105017
[hep-ph/9912274].
%%CITATION = HEP-PH 9912274;%%

%\cite{Boyanovsky:2001nt}
\bibitem{Boyanovsky:2001nt}
D.~Boyanovsky, H.~J.~de Vega and M.~Simionato,
%``Relaxing near the critical point,''
Phys.\ Rev.\ D {\bibbf 63} (2001) 045007
[hep-ph/0004159].
%%CITATION = HEP-PH 0004159;%%

%\cite{Zurek:1985qw}
\bibitem{Zurek:1985qw}
W.~H.~Zurek,
%``Cosmological Experiments In Superfluid Helium?,''
Nature {\bibbf 317} (1985) 505.
%%CITATION = NATUA,317,505;%%

\bibitem{Zinn-Justin}
J.~Zinn-Justin,
{\bibit Quantum Field Theory and Critical Phenomena}
(Clarendon Press, Oxford, 1989).

%\cite{Stephanov:1998dy}
\bibitem{Stephanov:1998dy}
M.~Stephanov, K.~Rajagopal and E.~Shuryak,
%``Signatures of the tricritical point in {QCD},''
Phys.\ Rev.\ Lett.\  {\bibbf 81} (1998) 4816
[hep-ph/9806219].
%%CITATION = HEP-PH 9806219;%%

\bibitem{TurokNature}
I.~Chuang, R.~Durrer, N.~Turok and B.~Yurke,
Science {\bibbf 251} (1991) 1336.

%\cite{Bowick:1994rz}
\bibitem{Bowick:1994rz}
M.~J.~Bowick, L.~Chandar, E.~A.~Schiff and A.~M.~Srivastava,
%``The Cosmological Kibble mechanism in the laboratory: String formation in liquid crystals,''
Science {\bibbf 263} (1994) 943
[hep-ph/9208233].
%%CITATION = HEP-PH 9208233;%%

\bibitem{Hendry}
P.~C.~Hendry {\bibit et al.},
Nature {\bibbf 368} (1995) 315.

\bibitem{Bauerle}
C.~B\"auerle {\bibit el al.},
Nature {\bibbf 382} (1996) 332.

%\cite{Ruutu:1996qz}
\bibitem{Ruutu:1996qz}
V.~M.~Ruutu {\bibit et al.},
%``Big bang simulation in superfluid 3He-B -- Vortex nucleation in neutron-irradiated superflow,''
Nature {\bibbf 382} (1996) 334
[cond-mat/9512117].
%%CITATION = COND-MAT 9512117;%%

\bibitem{Dodd}
M.~E.~Dodd {\bibit et al.},
Phys.~Rev.~Lett.~{\bibbf 81} (1998) 3703
[cond-mat/9808117].

%\cite{Hindmarsh:2000kd}
\bibitem{Hindmarsh:2000kd}
M.~Hindmarsh and A.~Rajantie,
%``Defect formation and local gauge invariance,''
Phys.\ Rev.\ Lett.\  {\bibbf 85} (2000) 4660
[cond-mat/0007361].
%%CITATION = COND-MAT 0007361;%%

%\cite{Hindmarsh:1995re}
\bibitem{Hindmarsh:1995re}
M.~B.~Hindmarsh and T.~W.~B.~Kibble,
%``Cosmic strings,''
Rept.\ Prog.\ Phys.\  {\bibbf 58} (1995) 477
[hep-ph/9411342].
%%CITATION = HEP-PH 9411342;%%

%\cite{Magueijo:2000se}
\bibitem{Magueijo:2000se}
J.~Magueijo and R.~H.~Brandenberger,
%``Cosmic defects and cosmology,''
in {\bibit
Large Scale Structure Formation}, 
edited by R.~Mansouri and R.~H.~Brandenberger (Kluwer, Dordrecht, 2000)
[astro-ph/0002030].
%%CITATION = ASTRO-PH 0002030;%%

%\cite{Nielsen:1973cs}
\bibitem{Nielsen:1973cs}
H.~B.~Nielsen and P.~Olesen,
%``Vortex Line Models For Dual Strings,''
Nucl.\ Phys.\ B {\bibbf 61} (1973) 45.
%%CITATION = NUPHA,B61,45;%%

%\cite{Jacobs:1979ch}
\bibitem{Jacobs:1979ch}
L.~Jacobs and C.~Rebbi,
%``Interaction Energy Of Superconducting Vortices,''
Phys.\ Rev.\ B {\bibbf 19} (1979) 4486.
%%CITATION = PHRVA,B19,4486;%%

\bibitem{Tilley}
D.~R.~Tilley and J.~Tilley, {\bibit Superfluidity and Superconductivity}
(IOP, Bristol, 1990).

%\cite{Bogomolny:1976de}
\bibitem{Bogomolny:1976de}
E.~B.~Bogomolny,
%``Stability Of Classical Solutions,''
Sov.\ J.\ Nucl.\ Phys.\  {\bibbf 24} (1976) 449
[Yad.\ Fiz.\  {\bibbf 24} (1976) 861].
%%CITATION = SJNCA,24,449.1976\ YAFIA,24,861;%%

%\cite{'tHooft:1974qc}
\bibitem{'tHooft:1974qc}
G.~'t Hooft,
%``Magnetic Monopoles In Unified Gauge Theories,''
Nucl.\ Phys.\ B {\bibbf 79} (1974) 276.
%%CITATION = NUPHA,B79,276;%%

%\cite{Polyakov:1974ek}
\bibitem{Polyakov:1974ek}
A.~M.~Polyakov,
%``Particle Spectrum In Quantum Field Theory,''
JETP Lett.\  {\bibbf 20} (1974) 194
[Pisma Zh.\ Eksp.\ Teor.\ Fiz.\  {\bibbf 20} (1974) 430].
%%CITATION = JTPLA,20,194;%%

%\cite{Preskill:1979zi}
\bibitem{Preskill:1979zi}
J.~P.~Preskill,
%``Cosmological Production Of Superheavy Magnetic Monopoles,''
Phys.\ Rev.\ Lett.\  {\bibbf 43} (1979) 1365.
%%CITATION = PRLTA,43,1365;%%

%\cite{Guth:1981zm}
\bibitem{Guth:1981zm}
A.~H.~Guth,
%``The Inflationary Universe: A Possible Solution To The Horizon And Flatness Problems,''
Phys.\ Rev.\ D {\bibbf 23} (1981) 347.
%%CITATION = PHRVA,D23,347;%%

\bibitem{Kapusta}
J.~I.~Kapusta,
{\bibit Finite-Temperature Field Theory} (Cambridge University Press,
Cambridge, 1989).

%\cite{Gross:1981br}
\bibitem{Gross:1981br}
D.~J.~Gross, R.~D.~Pisarski and L.~G.~Yaffe,
%``QCD And Instantons At Finite Temperature,''
Rev.\ Mod.\ Phys.\  {\bibbf 53} (1981) 43.
%%CITATION = RMPHA,53,43;%%

%\cite{Ginsparg:1980ef}
\bibitem{Ginsparg:1980ef}
P.~Ginsparg,
%``First Order And Second Order Phase Transitions In Gauge Theories At Finite Temperature,''
Nucl.\ Phys.\ B {\bibbf 170} (1980) 388.
%%CITATION = NUPHA,B170,388;%%

%\cite{Kajantie:1996dw}
\bibitem{Kajantie:1996dw}
K.~Kajantie, M.~Laine, K.~Rummukainen and M.~Shaposhnikov,
%``Generic rules for high temperature dimensional reduction and their application to the standard model,''
Nucl.\ Phys.\ B {\bibbf 458} (1996) 90
[hep-ph/9508379].
%%CITATION = HEP-PH 9508379;%%

%\cite{Kajantie:1997qd}
\bibitem{Kajantie:1997qd}
K.~Kajantie, M.~Laine, K.~Rummukainen and M.~Shaposhnikov,
%``A non-perturbative analysis of the finite T phase transition in  SU(2) x U(1) electroweak theory,''
Nucl.\ Phys.\ B {\bibbf 493} (1997) 413
[hep-lat/9612006].
%%CITATION = HEP-LAT 9612006;%%

%\cite{Jackiw:1974cv}
\bibitem{Jackiw:1974cv}
R.~Jackiw,
%``Functional Evaluation Of The Effective Potential,''
Phys.\ Rev.\ D {\bibbf 9} (1974) 1686.
%%CITATION = PHRVA,D9,1686;%%

%\cite{Elitzur:1975im}
\bibitem{Elitzur:1975im}
S.~Elitzur,
%``Impossibility Of Spontaneously Breaking Local Symmetries,''
Phys.\ Rev.\ D {\bibbf 12} (1975) 3978.
%%CITATION = PHRVA,D12,3978;%%

%\cite{Coleman:1973jx}
\bibitem{Coleman:1973jx}
S.~Coleman and E.~Weinberg,
%``Radiative Corrections As The Origin Of Spontaneous Symmetry Breaking,''
Phys.\ Rev.\ D {\bibbf 7} (1973) 1888.
%%CITATION = PHRVA,D7,1888;%%

\bibitem{HLM}
B.~I.~Halperin, T.~C.~Lubensky and S.-K. Ma,
Phys.~Rev.~Lett.~{\bibbf 32} (1974) 292.

%\cite{Fradkin:1979dv}
\bibitem{Fradkin:1979dv}
E.~Fradkin and S.~H.~Shenker,
%``Phase Diagrams Of Lattice Gauge Theories With Higgs Fields,''
Phys.\ Rev.\ D {\bibbf 19} (1979) 3682.
%%CITATION = PHRVA,D19,3682;%%

%\cite{Kleinert:1982dz}
\bibitem{Kleinert:1982dz}
H.~Kleinert,
%``Disorder Version Of The Abelian Higgs Model And The Order Of The Superconductive Phase Transition,''
Lett.\ Nuovo Cim.\  {\bibbf 35} (1982) 405.
%%CITATION = NCLTA,35,405;%%

%\cite{Kovner:1991pz}
\bibitem{Kovner:1991pz}
A.~Kovner, B.~Rosenstein and D.~Eliezer,
%``Photon as a Goldstone boson in (2+1)-dimensional Abelian gauge theories,''
Nucl.\ Phys.\ B {\bibbf 350} (1991) 325.
%%CITATION = NUPHA,B350,325;%%

%\cite{Langer:1967ax}
\bibitem{Langer:1967ax}
J.~S.~Langer,
%``Theory Of The Condensation Point,''
Annals Phys.\  {\bibbf 41} (1967) 108.
%%CITATION = APNYA,41,108;%%

%\cite{Langer:1969bc}
\bibitem{Langer:1969bc}
J.~S.~Langer,
%``Statistical Theory Of The Decay Of Metastable States,''
Annals Phys.\  {\bibbf 54} (1969) 258.
%%CITATION = APNYA,54,258;%%

%\cite{Coleman:1977py}
\bibitem{Coleman:1977py}
S.~Coleman,
%``The Fate Of The False Vacuum. 1. Semiclassical Theory,''
Phys.\ Rev.\ D {\bibbf 15} (1977) 2929
[Erratum-ibid.\ D {\bibbf 16} (1977) 1248].
%%CITATION = PHRVA,D15,2929;%%

%\cite{Moore:2001jw}
\bibitem{Moore:2001jw}
G.~D.~Moore and K.~Rummukainen,
%``Electroweak bubble nucleation, nonperturbatively,''
Phys.\ Rev.\ D {\bibbf 63} (2001) 045002
[hep-ph/0009132].
%%CITATION = HEP-PH 0009132;%%

%\cite{Kurki-Suonio:1996rk}
\bibitem{Kurki-Suonio:1996rk}
H.~Kurki-Suonio and M.~Laine,
%``Real-time history of the cosmological electroweak phase transition,''
Phys.\ Rev.\ Lett.\  {\bibbf 77} (1996) 3951
[hep-ph/9607382].
%%CITATION = HEP-PH 9607382;%%

%\cite{Ignatius:1994qn}
\bibitem{Ignatius:1994qn}
J.~Ignatius, K.~Kajantie, H.~Kurki-Suonio and M.~Laine,
%``The growth of bubbles in cosmological phase transitions,''
Phys.\ Rev.\ D {\bibbf 49} (1994) 3854
[astro-ph/9309059].
%%CITATION = ASTRO-PH 9309059;%%

%\cite{Zeldovich:1974uw}
\bibitem{Zeldovich:1974uw}
Y.~B.~Zeldovich, I.~Y.~Kobzarev and L.~B.~Okun,
%``Cosmological Consequences Of The Spontaneous Breakdown Of Discrete Symmetry,''
Zh.\ Eksp.\ Teor.\ Fiz.\  {\bibbf 67} (1974) 3
[Sov.\ Phys.\ JETP {\bibbf 40} (1974) 1].
%%CITATION = ZETFA,67,3;%%

%\cite{Vachaspati:1984dz}
\bibitem{Vachaspati:1984dz}
T.~Vachaspati and A.~Vilenkin,
%``Formation And Evolution Of Cosmic Strings,''
Phys.\ Rev.\ D {\bibbf 30} (1984) 2036.
%%CITATION = PHRVA,D30,2036;%%

%\cite{Prokopec:1991ab}
\bibitem{Prokopec:1991ab}
T.~Prokopec,
%``Formation of topological and nontopological defects in the early universe,''
Phys.\ Lett.\ B {\bibbf 262} (1991) 215.
%%CITATION = PHLTA,B262,215;%%

%\cite{Scherrer:1986sw}
\bibitem{Scherrer:1986sw}
R.~J.~Scherrer and J.~A.~Frieman,
%``Cosmic Strings As Random Walks,''
Phys.\ Rev.\ D {\bibbf 33} (1986) 3556.
%%CITATION = PHRVA,D33,3556;%%

%\cite{Borrill:1995gu}
\bibitem{Borrill:1995gu}
J.~Borrill, T.~W.~B.~Kibble, T.~Vachaspati and A.~Vilenkin,
%``Defect production in slow first order phase transitions,''
Phys.\ Rev.\ D {\bibbf 52} (1995) 1934
[hep-ph/9503223].
%%CITATION = HEP-PH 9503223;%%

%\cite{deLaix:1999xz}
\bibitem{deLaix:1999xz}
A.~A.~de Laix and T.~Vachaspati,
%``On random bubble lattices,''
Phys.\ Rev.\ D {\bibbf 59} (1999) 045017
[hep-ph/9802423].
%%CITATION = HEP-PH 9802423;%%

%\cite{Melfo:1995cv}
\bibitem{Melfo:1995cv}
A.~Melfo and L.~Perivolaropoulos,
%``Formation of vortices in first order phase transitions,''
Phys.\ Rev.\ D {\bibbf 52} (1995) 992
[hep-ph/9501284].
%%CITATION = HEP-PH 9501284;%%

%\cite{Ferrera:1996ef}
\bibitem{Ferrera:1996ef}
A.~Ferrera and A.~Melfo,
%``Bubble Collisions and Defect Formation in a Damping Environment,''
Phys.\ Rev.\ D {\bibbf 53} (1996) 6852
[hep-ph/9512290].
%%CITATION = HEP-PH 9512290;%%

%\cite{Rudaz:1993wy}
\bibitem{Rudaz:1993wy}
S.~Rudaz and A.~Mohan Srivastava,
%``On the production of flux vortices and magnetic monopoles in phase transitions,''
Mod.\ Phys.\ Lett.\ A {\bibbf 8} (1993) 1443
[hep-ph/9212279].
%%CITATION = HEP-PH 9212279;%%

%\cite{Hindmarsh:1994av}
\bibitem{Hindmarsh:1994av}
M.~Hindmarsh, A.~Davis and R.~Brandenberger,
%``Formation of topological defects in first order phase transitions,''
Phys.\ Rev.\ D {\bibbf 49} (1994) 1944
[hep-ph/9307203].
%%CITATION = HEP-PH 9307203;%%

%\cite{Kibble:1995aa}
\bibitem{Kibble:1995aa}
T.~W.~B.~Kibble and A.~Vilenkin,
%``Phase equilibration in bubble collisions,''
Phys.\ Rev.\ D {\bibbf 52} (1995) 679
[hep-ph/9501266].
%%CITATION = HEP-PH 9501266;%%

%\cite{Copeland:1996jz}
\bibitem{Copeland:1996jz}
E.~J.~Copeland and P.~M.~Saffin,
%``Bubble collisions in Abelian gauge theories and the geodesic rule,''
Phys.\ Rev.\ D {\bibbf 54} (1996) 6088
[hep-ph/9604231].
%%CITATION = HEP-PH 9604231;%%

%\cite{Kosterlitz:1973xp}
\bibitem{Kosterlitz:1973xp}
J.~M.~Kosterlitz and D.~J.~Thouless,
%``Ordering, Metastability And Phase Transitions In Two-Dimensional  Systems,''
J.\ Phys.\ C {\bibbf 6} (1973) 1181.
%%CITATION = JPCBA,C6,1181;%%

\bibitem{MerminWagner}
N.~C.~Mermin and H.~Wagner,
Phys.\ Rev.\ Lett.\ {\bibbf 17} (1966) 1133.

%\cite{Coleman:1973ci}
\bibitem{Coleman:1973ci}
S.~Coleman,
%``There Are No Goldstone Bosons In Two-Dimensions,''
Commun.\ Math.\ Phys.\  {\bibbf 31} (1973) 259.
%%CITATION = CMPHA,31,259;%%

\bibitem{KibbleInhomog}
T.~W.~B.~Kibble and G.~E.~Volovik, 
Pis'ma Zh. \'Eksp. Teor. Fiz. {\bibbf 65} (1997) 96
[JETP Lett. {\bibbf 65} (1997) 102].

\bibitem{Grisha2000}
G.~E.~Volovik,
Physica B {\bibbf 280} (2000) 122.

%\cite{Kibble:1980mv}
\bibitem{Kibble:1980mv}
T.~W.~B.~Kibble,
%``Some Implications Of A Cosmological Phase Transition,''
Phys.\ Rept.\  {\bibbf 67} (1980) 183.
%%CITATION = PRPLC,67,183;%%

%\cite{Zurek:1993ek}
\bibitem{Zurek:1993ek}
W.~H.~Zurek,
%``Cosmic strings in laboratory superfluids and the topological remnants of other phase transitions,''
Acta Phys.\ Polon.\ B {\bibbf 24} (1993) 1301.
%%CITATION = APPOA,B24,1301;%%

%\cite{Laguna:1998cf}
\bibitem{Laguna:1998cf}
P.~Laguna and W.~H.~Zurek,
%``Critical dynamics of symmetry breaking: Quenches, dissipation and  cosmology,''
Phys.\ Rev.\ D {\bibbf 58} (1998) 085021
[hep-ph/9711411].
%%CITATION = HEP-PH 9711411;%%

%\cite{Yates:1998kx}
\bibitem{Yates:1998kx}
A.~Yates and W.~H.~Zurek,
%``Vortex formation in two dimensions: When symmetry breaks, how big are  the pieces?,''
Phys.\ Rev.\ Lett.\  {\bibbf 80} (1998) 5477
[hep-ph/9801223].
%%CITATION = HEP-PH 9801223;%%

\bibitem{StephensNew}
G.~J.~Stephens, L.~M.~A.~Bettencourt and W.~H.~Zurek,
cond-mat/0108127.

%\cite{Zurek:1996sj}
\bibitem{Zurek:1996sj}
W.~H.~Zurek,
%``Cosmological Experiments in Condensed Matter Systems,''
Phys.\ Rept.\  {\bibbf 276} (1996) 177
[cond-mat/9607135].
%%CITATION = COND-MAT 9607135;%%

%\cite{Hindmarsh:2001vp}
\bibitem{Hindmarsh:2001vp}
M.~Hindmarsh and A.~Rajantie,
%``Phase transition dynamics in the hot Abelian Higgs model,''
Phys.\ Rev.\ D {\bibbf 64} (2001) 065016
[hep-ph/0103311].
%%CITATION = HEP-PH 0103311;%%

%\cite{Rajantie:2001na}
\bibitem{Rajantie:2001na}
A.~Rajantie,
%``Local gauge invariance and formation of topological defects,''
J.\ Low.\ Temp.\ Phys.\  {\bibbf 124} (2001) 5
[cond-mat/0102403].
%%CITATION = COND-MAT 0102403;%%

%\cite{Digal:1998ak}
\bibitem{Digal:1998ak}
S.~Digal, R.~Ray and A.~M.~Srivastava,
%``Observing Correlated Production of Defects-Antidefects in Liquid Crystals,''
Phys.\ Rev.\ Lett.\ {\bibbf 83} (1999) 5030.
[hep-ph/9805502].
%%CITATION = HEP-PH 9805502;%%

%\cite{Gill:1995ye}
\bibitem{Gill:1995ye}
A.~J.~Gill and R.~J.~Rivers,
%``The Dynamics of vortex and monopole production by quench induced phase separation,''
Phys.\ Rev.\ D {\bibbf 51} (1995) 6949
[hep-th/9410159].
%%CITATION = HEP-TH 9410159;%%

%\cite{Lombardo:2001vs}
\bibitem{Lombardo:2001vs}
F.~C.~Lombardo, F.~D.~Mazzitelli and R.~J.~Rivers,
%``Decoherence after a phase transition,''
hep-ph/0102152.
%%CITATION = HEP-PH 0102152;%%

\bibitem{Halperin}
B.~Halperin, 
in {\bibit Physics of Defects}, edited by R.~Balian, M.~Kleman,
and J.~P.~Poirier (North-Holland, New York, 1981).

\bibitem{Mazenko-Liu}
F.~Liu and G.~F.~Mazenko, Phys.\ Rev.\ B {\bibbf 46} (1992) 5963.

%\cite{Karra:1996xf}
\bibitem{Karra:1996xf}
G.~Karra and R.~J.~Rivers,
%``The Densities, Correlations and Length Distributions of Vortices Produced at a Gaussian Quench,''
hep-ph/9603413.
%%CITATION = HEP-PH 9603413;%%

%\cite{Karra:1997it}
\bibitem{Karra:1997it}
G.~Karra and R.~J.~Rivers,
%``Initial vortex densities after a temperature quench,''
Phys.\ Lett.\ B {\bibbf 414} (1997) 28
[hep-ph/9705243].
%%CITATION = HEP-PH 9705243;%%

%\cite{Karra:1998gn}
\bibitem{Karra:1998gn}
G.~Karra and R.~J.~Rivers,
%``A re-examination of quenches in He-4,''
Phys.\ Rev.\ Lett.\  {\bibbf 81} (1998) 3707
[hep-ph/9804206].
%%CITATION = HEP-PH 9804206;%%

\bibitem{Lythe1996}
G.~D.~Lythe,
Phys.\ Rev.\ E {\bibbf 53} (1996) R4271.

\bibitem{Moro1999}
E.~Moro and G.~Lythe,
Phys.\ Rev.\ E {\bibbf 59} (1999) R1303.

\bibitem{Dziarmaga1998}
J.~Dziarmaga,
Phys.\ Rev.\ Lett.\ {\bibbf 81} (1998) 1551.

%\cite{Cornwall:1974vz}
\bibitem{Cornwall:1974vz}
J.~M.~Cornwall, R.~Jackiw and E.~Tomboulis,
%``Effective Action For Composite Operators,''
Phys.\ Rev.\ D {\bibbf 10} (1974) 2428.
%%CITATION = PHRVA,D10,2428;%%

%\cite{Calzetta:1988cq}
\bibitem{Calzetta:1988cq}
E.~Calzetta and B.~L.~Hu,
%``Nonequilibrium Quantum Fields: Closed Time Path Effective Action, Wigner Function And Boltzmann Equation,''
Phys.\ Rev.\ D {\bibbf 37} (1988) 2878.
%%CITATION = PHRVA,D37,2878;%%

%\cite{Chang:1975dt}
\bibitem{Chang:1975dt}
S.~Chang,
%``Quantum Fluctuations In A Phi**4 Field Theory. 1. The Stability Of The Vacuum,''
Phys.\ Rev.\ D {\bibbf 12} (1975) 1071.
%%CITATION = PHRVA,D12,1071;%%

%\cite{Boyanovsky:1993pf}
\bibitem{Boyanovsky:1993pf}
D.~Boyanovsky, D.~Lee and A.~Singh,
%``Phase transitions out-of-equilibrium: Domain formation and growth,''
Phys.\ Rev.\ D {\bibbf 48} (1993) 800
[hep-th/9212083].
%%CITATION = HEP-TH 9212083;%%

%\cite{Antunes:1997na}
\bibitem{Antunes:1997na}
N.~D.~Antunes and L.~M.~Bettencourt,
%``Out of equilibrium dynamics of a quench-induced phase transition and  topological defect formation,''
Phys.\ Rev.\ D {\bibbf 55} (1997) 925
[hep-ph/9605277].
%%CITATION = HEP-PH 9605277;%%

%\cite{Bowick:1998kd}
\bibitem{Bowick:1998kd}
M.~Bowick and A.~Momen,
%``Domain formation in finite-time quenches,''
Phys.\ Rev.\ D {\bibbf 58} (1998) 085014
[hep-ph/9803284].
%%CITATION = HEP-PH 9803284;%%

%\cite{Stephens:1999sm}
\bibitem{Stephens:1999sm}
G.~J.~Stephens, E.~A.~Calzetta, B.~L.~Hu and S.~A.~Ramsey,
%``Defect formation and critical dynamics in the early universe,''
Phys.\ Rev.\ D {\bibbf 59} (1999) 045009
[gr-qc/9808059].
%%CITATION = GR-QC 9808059;%%

%\cite{Berges:2001ur}
\bibitem{Berges:2001ur}
J.~Berges and J.~Cox,
%``Thermalization of quantum fields from time-reversal invariant evolution  equations,''
Phys.\ Lett.\ B {\bibbf 517} (2001) 369
[hep-ph/0006160].
%%CITATION = HEP-PH 0006160;%%

%\cite{Aarts:2001qa}
\bibitem{Aarts:2001qa}
G.~Aarts and J.~Berges,
%``Nonequilibrium time evolution of the spectral function in quantum field  theory,''
hep-ph/0103049.
%%CITATION = HEP-PH 0103049;%%

%\cite{Aarts:2001yn}
\bibitem{Aarts:2001yn}
G.~Aarts and J.~Berges,
%``Classical aspects of quantum fields far from equilibrium,''
hep-ph/0107129.
%%CITATION = HEP-PH 0107129;%%

%\cite{Cheetham:1996nd}
\bibitem{Cheetham:1996nd}
G.~J.~Cheetham and E.~J.~Copeland,
%``Quantum dynamics beyond the Gaussian approximation,''
Phys.\ Rev.\ D {\bibbf 53} (1996) 4125
[gr-qc/9503043].
%%CITATION = GR-QC 9503043;%%

%\cite{Kim:2000xd}
\bibitem{Kim:2000xd}
S.~P.~Kim and F.~C.~Khanna,
%``Non-Gaussian effects on domain growth,''
hep-ph/0011115.
%%CITATION = HEP-PH 0011115;%%

%\cite{Salle:2001hd}
\bibitem{Salle:2001hd}
M.~Salle, J.~Smit and J.~C.~Vink,
%``Thermalization in a Hartree ensemble approximation to quantum field  dynamics,''
Phys.\ Rev.\ D {\bibbf 64} (2001) 025016
[hep-ph/0012346].
%%CITATION = HEP-PH 0012346;%%

%\cite{Grigoriev:1988bd}
\bibitem{Grigoriev:1988bd}
D.~Y.~Grigoriev and V.~A.~Rubakov,
%``Soliton Pair Creation At Finite Temperatures. Numerical Study In (1+1)-Dimensions,''
Nucl.\ Phys.\ B {\bibbf 299} (1988) 67.
%%CITATION = NUPHA,B299,67;%%

\bibitem{Montvay}
I.~Montvay and G.~M\"unster,
{\bibit Quantum Fields of a Lattice} (Cambridge University Press,
Cambridge, 1997).

\bibitem{Rothe}
H.~J.~Rothe,
{\bibit Lattice Gauge Theories: an Introduction}
(World Scientific, Singapore, 1998).

%\cite{Parisi:1981ys}
\bibitem{Parisi:1981ys}
G.~Parisi and Y.~Wu,
%``Perturbation Theory Without Gauge Fixing,''
Sci.\ Sin.\  {\bibbf 24} (1981) 483.
%%CITATION = SSINA,24,483;%%

%\cite{Ukawa:1985hr}
\bibitem{Ukawa:1985hr}
A.~Ukawa and M.~Fukugita,
%``Langevin Simulation Including Dynamical Quark Loops,''
Phys.\ Rev.\ Lett.\  {\bibbf 55} (1985) 1854.
%%CITATION = PRLTA,55,1854;%%

%\cite{Metropolis:1953am}
\bibitem{Metropolis:1953am}
N.~Metropolis, A.~W.~Rosenbluth, M.~N.~Rosenbluth, A.~H.~Teller and E.~Teller,
%``Equation Of State Calculations By Fast Computing Machines,''
J.\ Chem.\ Phys.\  {\bibbf 21} (1953) 1087.
%%CITATION = JCPSA,21,1087;%%

%\cite{Creutz:1980zw}
\bibitem{Creutz:1980zw}
M.~Creutz,
%``Monte Carlo Study Of Quantized SU(2) Gauge Theory,''
Phys.\ Rev.\ D {\bibbf 21} (1980) 2308.
%%CITATION = PHRVA,D21,2308;%%

%\cite{Duane:1987de}
\bibitem{Duane:1987de}
S.~Duane, A.~D.~Kennedy, B.~J.~Pendleton and D.~Roweth,
%``Hybrid Monte Carlo,''
Phys.\ Lett.\ B {\bibbf 195} (1987) 216.
%%CITATION = PHLTA,B195,216;%%

%\cite{Laine:1995ag}
\bibitem{Laine:1995ag}
M.~Laine,
%``Exact relation of lattice and continuum parameters in three-dimensional SU(2) + Higgs theories,''
Nucl.\ Phys.\ B {\bibbf 451} (1995) 484
[hep-lat/9504001].
%%CITATION = HEP-LAT 9504001;%%

%\cite{Laine:1998dy}
\bibitem{Laine:1998dy}
M.~Laine and A.~Rajantie,
%``Lattice-continuum relations for 3d SU(N)+Higgs theories,''
Nucl.\ Phys.\ B {\bibbf 513} (1998) 471
[hep-lat/9705003].
%%CITATION = HEP-LAT 9705003;%%

%\cite{Bodeker:1995pp}
\bibitem{Bodeker:1995pp}
D.~Bodeker, L.~McLerran and A.~Smilga,
%``Really computing nonperturbative real time correlation functions,''
Phys.\ Rev.\ D {\bibbf 52} (1995) 4675
[hep-th/9504123].
%%CITATION = HEP-TH 9504123;%%


%\cite{Moore:2000fs}
\bibitem{Moore:2000fs}
G.~D.~Moore and K.~Rummukainen,
%``Classical sphaleron rate on fine lattices,''
Phys.\ Rev.\ D {\bibbf 61} (2000) 105008
[hep-ph/9906259].
%%CITATION = HEP-PH 9906259;%%

%\cite{Laguna:1997pv}
\bibitem{Laguna:1997pv}
P.~Laguna and W.~H.~Zurek,
%``Density of kinks after a quench: When symmetry breaks, how big are the  pieces?,''
Phys.\ Rev.\ Lett.\  {\bibbf 78} (1997) 2519
[gr-qc/9607041].
%%CITATION = GR-QC 9607041;%%

%\cite{Laguna:1997sm}
\bibitem{Laguna:1997sm}
P.~Laguna and W.~H.~Zurek,
%``Density of topological defects after a quench,''
cond-mat/9705141.
%%CITATION = COND-MAT 9705141;%%


%\cite{Antunes:1999rz}
\bibitem{Antunes:1999rz}
N.~D.~Antunes, L.~M.~Bettencourt and W.~H.~Zurek,
%``Vortex string formation in a 3D U(1) temperature quench,''
Phys.\ Rev.\ Lett.\  {\bibbf 82} (1999) 2824
[hep-ph/9811426].
%%CITATION = HEP-PH 9811426;%%

%\cite{Bettencourt:2000jv}
\bibitem{Bettencourt:2000jv}
L.~M.~Bettencourt, N.~D.~Antunes and W.~H.~Zurek,
%``The Ginzburg regime and its effects on topological defect formation,''
Phys.\ Rev.\ D {\bibbf 62} (2000) 065005
[hep-ph/0001205].
%%CITATION = HEP-PH 0001205;%%

%\cite{Bowick:2001xg}
\bibitem{Bowick:2001xg}
M.~J.~Bowick, A.~Cacciuto and A.~Travesset,
%``The formation of vortex loops (strings) in continuous phase  transitions,''
cond-mat/0107188.
%%CITATION = COND-MAT 0107188;%%

%\cite{Stephens:2000qv}
\bibitem{Stephens:2000qv}
G.~J.~Stephens,
%``Unraveling critical dynamics: The formation and evolution of  topological textures,''
Phys.\ Rev.\ D {\bibbf 61} (2000) 085002
[hep-ph/9911247].
%%CITATION = HEP-PH 9911247;%%

%\cite{Ambjorn:1995xm}
\bibitem{Ambjorn:1995xm}
J.~Ambjorn and A.~Krasnitz,
%``The classical sphaleron transition rate exists and is equal to $1.1(\alpha_w T)~4$,''
Phys.\ Lett.\ B {\bibbf 362} (1995) 97
[hep-ph/9508202].
%%CITATION = HEP-PH 9508202;%%

%\cite{Moriarty:1988fx}
\bibitem{Moriarty:1988fx}
K.~J.~Moriarty, E.~Myers and C.~Rebbi,
%``Dynamical Interactions Of Flux Vortices In Superconductors,''
Phys.\ Lett.\ B {\bibbf 207} (1988) 411.
%%CITATION = PHLTA,B207,411;%%

%\cite{Kajantie:1998bg}
\bibitem{Kajantie:1998bg}
K.~Kajantie, M.~Karjalainen, M.~Laine, J.~Peisa and A.~Rajantie,
%``Thermodynamics of gauge-invariant U(1) vortices from lattice Monte  Carlo simulations,''
Phys.\ Lett.\ B {\bibbf 428} (1998) 334
[hep-ph/9803367].
%%CITATION = HEP-PH 9803367;%%

%\cite{Ranft:1983hf}
\bibitem{Ranft:1983hf}
J.~Ranft, J.~Kripfganz and G.~Ranft,
%``Phase Structure, Magnetic Monopoles And Vortices In The Lattice Abelian Higgs Model,''
Phys.\ Rev.\ D {\bibbf 28} (1983) 360.
%%CITATION = PHRVA,D28,360;%%

%\cite{Ibaceta:1999yy}
\bibitem{Ibaceta:1999yy}
D.~Ibaceta and E.~Calzetta,
%``Counting defects in an instantaneous quench,''
Phys.\ Rev.\  {\bibbf E60} (1999) 2999
[hep-ph/9810301].
%%CITATION = HEP-PH 9810301;%%

%\cite{Dziarmaga:1998ie}
\bibitem{Dziarmaga:1998ie}
J.~Dziarmaga, P.~Laguna and W.~H.~Zurek,
%``Symmetry breaking with a slant: Topological defects after an inhomogeneous quench,''
Phys.\ Rev.\ Lett.\ {\bibbf 82} (1999) 4749
[cond-mat/9810396].
%%CITATION = COND-MAT 9810396;%%

\bibitem{Aranson1999}
I.~S.~Aranson, N.~B.~Kopnin and V.~M.~Vinokur, 
Phys.\ Rev.\ Lett.\ {\bibbf 83} (1999) 2600.

\bibitem{Aranson2001}
I.~S.~Aranson, N.~B.~Kopnin and V.~M.~Vinokur, 
Phys.\ Rev.\ B {\bibbf 63} (2001) 184501. 

%\cite{Aarts:1997qi}
\bibitem{Aarts:1997qi}
G.~Aarts and J.~Smit,
%``Finiteness of hot classical scalar field theory and the plasmon  damping rate,''
Phys.\ Lett.\ B {\bibbf 393} (1997) 395
[hep-ph/9610415].
%%CITATION = HEP-PH 9610415;%%



%\cite{Pisarski:1989vd}
\bibitem{Pisarski:1989vd}
R.~D.~Pisarski,
%``Scattering Amplitudes In Hot Gauge Theories,''
Phys.\ Rev.\ Lett.\  {\bibbf 63} (1989) 1129.
%%CITATION = PRLTA,63,1129;%%

%\cite{Kraemmer:1995az}
\bibitem{Kraemmer:1995az}
U.~Kraemmer, A.~K.~Rebhan and H.~Schulz,
%``Resummations in hot scalar electrodynamics,''
Annals Phys.\  {\bibbf 238}, 286 (1995)
[hep-ph/9403301].
%%CITATION = HEP-PH 9403301;%%

%\cite{Nauta:2000cm}
\bibitem{Nauta:2000cm}
B.~J.~Nauta,
%``Counterterms for linear divergences in real-time classical gauge  theories at high temperature,''
Nucl.\ Phys.\ B {\bibbf 575} (2000) 383
[hep-ph/9906389].
%%CITATION = HEP-PH 9906389;%%

%\cite{Rajantie:1999mp}
\bibitem{Rajantie:1999mp}
A.~Rajantie and M.~Hindmarsh,
%``Simulating hot Abelian gauge dynamics,''
Phys.\ Rev.\ D {\bibbf 60} (1999) 096001
[hep-ph/9904270].
%%CITATION = HEP-PH 9904270;%%

%\cite{Moore:1998sn}
\bibitem{Moore:1998sn}
G.~D.~Moore, C.~Hu and B.~Muller,
%``Chern-Simons number diffusion with hard thermal loops,''
Phys.\ Rev.\ D {\bibbf 58} (1998) 045001
[hep-ph/9710436].
%%CITATION = HEP-PH 9710436;%%

%\cite{Blaizot:1999xk}
\bibitem{Blaizot:1999xk}
J.~Blaizot and E.~Iancu,
%``A Boltzmann equation for the {QCD} plasma,''
Nucl.\ Phys.\ B {\bibbf 557} (1999) 183
[hep-ph/9903389].
%%CITATION = HEP-PH 9903389;%%

%\cite{Bodeker:2000gx}
\bibitem{Bodeker:2000gx}
D.~Bodeker, G.~D.~Moore and K.~Rummukainen,
%``Chern-Simons number diffusion and hard thermal loops on the lattice,''
Phys.\ Rev.\ D {\bibbf 61} (2000) 056003
[hep-ph/9907545].
%%CITATION = HEP-PH 9907545;%%

\bibitem{Ducci1999}
S.~Ducci, P.~L.~Ramazza, W.~Gonzalez-Vinas and F.~T.~Arecchi,
Phys.\ Rev.\ Lett.\ {\bibbf 83} (1999) 5210.

\bibitem{Casado2001}
S.~Casado, W.~González-Viñas, H.~Mancini and S.~Boccaletti,
Phys.\ Rev.\ E {\bibbf 63} (2001) 057301.

%\cite{Anglin:1999pm}
\bibitem{Anglin:1999pm}
J.~R.~Anglin and W.~H.~Zurek,
%``Winding up by a quench: vortices in the wake of rapid Bose-Einstein condensation,''
Phys.\ Rev.\ Lett.\  {\bibbf 83} (1999) 1707
[quant-ph/9804035].
%%CITATION = QUANT-PH 9804035;%%

\bibitem{Carmi}
R.~Carmi and E.~Polturak,
Phys.\ Rev.\ B {\bibbf 60} (1999) 7595.

\bibitem{Carmi_Jos}
R.~Carmi, E.~Polturak and G.~Koren,
Phys.\ Rev.\ Lett.\ {\bibbf 84} (2000) 4966.

\bibitem{deGennes}
P.~G.~de~Gennes and J.~Prost, {\bibit The Physics of Liquid Crystals} 
(Clarendon, Oxford, 1993).

\bibitem{GinzPit}
V.~L.~Ginzburg and L.~P.~Pitaesvkii,
Zh.\ Eksp.\ Teor.\ Fiz.\ {\bibbf 34} (1958) 1240
[Sov.\ Phys.\ JETP {\bibbf 7} (1958) 858].

\bibitem{Gill1996}
A.~J.~Gill and T.~W.~B.~Kibble,
J.\ Phys.\ A {\bibbf 29} (1996) 4289.

\bibitem{Dodd2}
M.E.~Dodd {\bibit et al.},
J.~Low~Temp.~Phys.~{\bibbf 115} (1999) 89.
[cond-mat/9810107].

\bibitem{VolovikMineev}
G.~E.~Volovin and V.~P.~Mineev,
Sov.\ Phys.\ JETP {\bibbf 45} (1977) 2256
[Zh.\ Eksp.\ Teor.\ Fiz.\ {\bibbf 72} (1977) 1186].

%\cite{Eltsov:2000ke}
\bibitem{Eltsov:2000ke}
V.~B.~Eltsov, T.~W.~B.~Kibble, M.~Krusius, V.~M.~Ruutu and G.~E.~Volovik,
%``Composite defect extends cosmology: He-3 analogy,''
Phys.\ Rev.\ Lett.\  {\bibbf 85} (2000) 4739
[cond-mat/0007369].
%%CITATION = COND-MAT 0007369;%%

%\cite{Rudaz:1999ra}
\bibitem{Rudaz:1999ra}
S.~Rudaz, A.~M.~Srivastava and S.~Varma,
%``Probing gauge string formation in a superconducting phase transition,''
Int.\ J.\ Mod.\ Phys.\ A {\bibbf 14} (1999) 1591.
%%CITATION = IMPAE,A14,1591;%%

%\cite{Zurek:2000ym}
\bibitem{Zurek:2000ym}
W.~H.~Zurek, L.~M.~Bettencourt, J.~Dziarmaga and N.~D.~Antunes,
%``Shards of broken symmetry: Topological defects as traces of the phase transition dynamics,''
Acta Phys.\ Polon.\ B {\bibbf 31} (2000) 2937.
%%CITATION = APPOA,B31,2937;%%

\bibitem{Bending}
S.~J.~Bending,
Adv.~Phys.~{\bibbf 48} (1999) 449.

\bibitem{Harada}
K.~Harada {\bibit et al.},
Nature {\bibbf 360} (1992) 51.

\bibitem{Goa}
P.~E.~Goa, H.~Hauglin, M.~Baziljevich, E.~Il'yashenko, P.~L.~Gammel
and T.~H.~Johansen,
cond-mat/0104280.

\bibitem{Moshchalkov}
V.~V.~Moshchalkov {\bibit et al.}
Phys.\ Rev.\ B {\bibbf57} (1998) 3615.

%\cite{Kavoussanaki:2000tj}
\bibitem{Kavoussanaki:2000tj}
E.~Kavoussanaki, R.~Monaco and R.~J.~Rivers,
%``Testing the Kibble-Zurek Scenario with Annular Josephson Tunnel Junctions,''
Phys.\ Rev.\ Lett.\  {\bibbf 85} (2000) 3452
[cond-mat/0005145].
%%CITATION = COND-MAT 0005145;%%









\end{thebibliography}
\end{document}